\documentclass{bmcart}

\usepackage[utf8]{inputenc} 

\usepackage{booktabs} 
\usepackage{amsmath}
\usepackage{graphicx}
\usepackage{multirow}
\usepackage{array}
\usepackage{color}
\usepackage{balance} 
\usepackage[absolute]{textpos} 
\usepackage{epstopdf}
\usepackage[parfill]{parskip}
\usepackage{algorithm}
\usepackage[noend]{algpseudocode}
\makeatletter
\def\BState{\State\hskip-\ALG@thistlm}
\makeatother

\usepackage{amssymb}

\algdef{SE}[SUBALG]{Indent}{EndIndent}{}{\algorithmicend\ }%
\algtext*{Indent}
\algtext*{EndIndent}

\algdef{SE}[FUNCTION]{Function}{EndFunction}%
[2]{\algorithmicfunction\ \textproc{#1}\ifthenelse{\equal{#2}{}}{}{(#2)}}%
{\algorithmicend\ \algorithmicfunction}%

\newcommand\Algphase[1]{%
	\vspace*{-.7\baselineskip}\Statex\hspace*{\dimexpr-\algorithmicindent-2pt\relax}\rule{\linewidth}{0.4pt}%
	\Statex\hspace*{-\algorithmicindent}\textbf{#1}%
	\vspace*{-.7\baselineskip}\Statex\hspace*{\dimexpr-\algorithmicindent-2pt\relax}\rule{\linewidth}{0.4pt}%
}

\startlocaldefs
\endlocaldefs

\begin{document}

\begin{frontmatter}

\begin{fmbox}
\dochead{Research}


\title{Real-time Spatio-temporal Event Detection on Geotagged Social Media}


\author[
   addressref={aff1,aff3},
   email={georgey@ibm.com}
]{\inits{YG}\fnm{Yasmeen} \snm{George}}
\author[
   addressref={aff1},                   
   corref={aff1},                       
   email={karus@unimelb.edu.au}   
]{\inits{SK}\fnm{Shanika} \snm{Karunasekera}}
\author[
   addressref={aff1},
   email={aharwood@unimelb.edu.au}
]{\inits{AH}\fnm{Aaron} \snm{Harwood}}
\author[
   addressref={aff2},
   email={kwanhui_lim@sutd.edu.sg}
]{\inits{KHL}\fnm{Kwan Hui} \snm{Lim}}


\address[id=aff1]{
  \orgname{School of Computing and Information Systems, The University of Melbourne}, 
  \cny{Australia}                                    
}
\address[id=aff2]{%
  \orgname{Information Systems Technology and Design Pillar, Singapore University of Technology and Design},
  \cny{Singapore}
}
\address[id=aff3]{
  \orgname{IBM Research - Australia}, 
  \cny{Australia}                                    
}


\begin{artnotes}
\end{artnotes}

\end{fmbox}


\begin{abstractbox}

\begin{abstract} 
A key challenge in mining social media data streams is to identify events which are actively discussed by a group of people in a specific local or global area. Such events are useful for early warning for accident, protest, election or breaking news. However, neither the list of events nor the resolution of both event time and space is fixed or known beforehand. In this work, we propose an online spatio-temporal event detection system using social media that is able to detect events at different time and space resolutions. First, to address the challenge related to the unknown spatial resolution of events, a quad-tree method is exploited in order to split the geographical space into multiscale regions based on the density of social media data. Then, a statistical unsupervised approach is performed that involves Poisson distribution and a smoothing method for highlighting regions with unexpected density of social posts. Further, event duration is precisely estimated by merging events happening in the same region at consecutive time intervals. A post processing stage is introduced to filter out events that are spam, fake or wrong. Finally, we incorporate simple semantics by using social media entities 
to assess the integrity, and accuracy of detected events. The proposed method is evaluated using different social media datasets: Twitter and Flickr for different cities: Melbourne, London, Paris and New York. To verify the effectiveness of the proposed method, we compare our results with two baseline algorithms based on fixed split of geographical space and clustering method. For performance evaluation, we manually compute recall and precision. We also propose a new quality measure named strength index, which automatically measures how accurate the reported event is.
\end{abstract}




\end{abstractbox}
%

\end{frontmatter}




\section{Introduction}

With the ubiquitous nature of smart phones, Twitter and other social media services, are frequently used as a source of news and other information~\cite{wang2020dialogic,petrovic2013can,newman2011mainstream}. In addition, people use such social media to share news and photos about various events they may encounter in their daily lives, oftentimes in real-time as these events unfold. Due to real-time sharing by people, social media serves as an efficient source of breaking news compared to traditional media, which are either slow to pick up such information or do not give a complete and accurate picture of the news and events. Due to these reasons, many researchers and organizations are relying on social media for obtaining timely news. One emerging use case of significant importance is where social media information is used for real-time event detection. For example, governments and organizations may be interested in events that are occurring in a particular geographical area, such as detecting a bush fire near residential areas, traffic congestion and accidents on highways, protests and other security incidents in the city. Being able to promptly detect such events is important as this early detection allows the relevant authorities and organizations to make the necessary responses to address these potentially adversarial events.

The traditional approach to detect events in social streams is to track the aggregate trend changes based on the count of geotagged social media data at given location and time. This approach is very closely related to topic detection and tracking, where an event is conventionally represented by a number of keywords, topics or tweets showing bursts in appearance count, i.e. keywords that are mentioned significantly more often during a (not too short) time period than in the period preceding it  \cite{doi:10.1137/1.9781611972825.54,popovici2014line}. Most of these existing  approaches detect events at fixed spatial and temporal resolutions, e.g., grids, which do not adequately capture the dynamic changes in tweeting volume across different areas and time \cite{zhang2016geoburst}. However, real-life events can occur at any spatial or temporal resolution, which is not known a priori and, therefore, algorithms that have fixed resolution result in suboptimal performance. While there are some approaches that are designed for detecting events at multiscale spatio-temporal resolutions \cite{walther2013geo,dong2015multiscale,Joan-C-2016}, they are essentially batch-based algorithms which are not directly applicable in the online real-time event detection scenarios.  While there are a few online spatial event detection algorithms proposed in the literature \cite{Andrienko-2015,wang2013real,hasan2018real}, they are fixed in terms of spatial and temporal resolution. Many of these works also utilized a supervised approach to event detection which may not work well for new types of events. 
In this paper, we aim to address the following problem: Given a stream of geotagged social media posts, how can we identify a set of posts that corresponds to a spatio-temporal event based on spatial and temporal proximity.

{\bf Research Objectives and Contributions}. 
Our main objective is to detect spatio-temporal events from social media, which could be any event that is being discussed loudly (frequently) in a specific local or global area. There are various challenges with this type of event detection. The first challenge associated with event detection is that there is no consensus among researchers on the definition of an event. The second challenge is that the location, time and the scale of the events (both in time and space) are not known before hand. Furthermore, the characteristics of past events may not be indicative of future events. Finally, the event detection algorithms developed need to be single pass and computationally efficient as we are interested in detecting events in real-time from high velocity data streams.

To address these challenges, we propose a novel approach to online spatio-temporal event detection that utilizes: (i) a quad-tree and Poisson model variant to dynamically identify events across different spatial scales; and (ii) a smoothing and filtering approach to effectively detect events with different temporal resolutions. The contribution of this paper can be summarized as follows~\footnote{This paper is an extended version of an earlier conference version~\cite{George-BigData19}, with additional content that includes: (i) more detailed description of our proposed algorithm with pseudo codes; (ii) additional experiments for the cities of London, Paris and New York; (iii) analysis of the effects of different parameter values; (iv) more in-depth discussion of our experiment results and findings.}: 

\begin{itemize}
	\item We leverage the quad-tree data structure for multi-scale event detection, to overcome the problem of detecting events with varying spatial coverage.
	\item We combine a Poisson model with a smoothing function for unsupervised event detection, to enable us to detect previously unseen events with different temporal resolutions. 
	\item We proposed a new event validation measure, strength index (SI), which automatically assesses the accuracy of the detected events by using social media entities. 
	\item We performed quantitative and comparative evaluations, which confirm the effectiveness of the proposed method in detecting new events correctly and completely.
	\item We demonstrated the generalizability of the proposed method by evaluating on different social media datasets, namely: Twitter and Flickr.
\end{itemize}

{\bf Structure and Organization}. This paper is structured as follows. Section~\ref{sect:relatedWork} reviews related work in the area of event detection, while Section~\ref{sectBgProblem} introduces the formulation of the event detection problem.
Section~\ref{sectAlgorithm} describes our proposed algorithm for location-based event detection, and Section~\ref{sectExperiments} shows the experimental results of our proposed algorithm against various baselines.
Finally, Section~\ref{sectConclusion} summarizes and concludes this paper.

\section{Related Work}
\label{sect:relatedWork}

In this section, we discuss various works that related to event detection on social media, ranging from general event detection works to those that focus on location-specific event detection.

\subsection{General Event Detection}

There are various related work that study the general problem of event detection using social media, without an explicit focus on the spatial aspect of events. These works aim to detect events on social media in the form of trending events based on posting patterns or specific pre-determined events. For example, numerous researchers have examined the problem of identifying trending and bursty events~\cite{cui-cikm12,Zubiaga-tkde16,xie-tkde16}, and detecting controversial events~\cite{popescu-cikm10, Dori-cikm13}. 

Among these works, there are many research that utilize supervised approaches for detecting events, such as~\cite{Zubiaga-tkde16} that used a Support Vector Machine for classifying tweets into one of the four topics of news, ongoing events, memes, and commemoratives. Others like Sakaki et al.~\cite{Sakaki-www10,Sakaki-tkde13} also use a trained Support Vector Machine to first determine if tweets are earthquake-related or not, then applying Kalman filtering and particle filtering on tweets to estimate the centres of these detected earthquakes. Popescu and Pennacchiotti~\cite{popescu-cikm10} proposed a Gradient Boosted Decision Tree trained on textual, social and news related features to determine if a set of tweets are controversy related or not. These works typically require a labelled dataset on which social media posts is associated with a specific event type, which works well for those events but may be challenging to generalize to unseen or new types of events. 

Another group of these works utilize largely unsupervised approaches for detecting events, without the need for explicit labels. For example, Weng and Lee~\cite{Weng-icwsm11} utilize (tweet) word signals derived from wavelet analysis, which are clustered together using a modularity-based graph partitioning to represent detected events. In a similar spirit, Aggarwal and Subbian~\cite{doi:10.1137/1.9781611972825.54} proposed an online clustering approach for detecting events based on the textual content of social media posts, their temporal distribution and the interaction network among users. Many of these works utilize clustering techniques or similar approaches for detecting events. However, many of these works aim to detect events without considering the spatial aspects of these events. Considering the spatial aspect of event is important for applications such as disaster detection and crisis management.

Similarly, there are also various web and mobile applications for tracking general events or retrieving tweets related to specific events~\cite{Li-icde12,lim-ecml18,Kwan-IUI21,Li-cikm12}. For a more detailed survey on general event detection, we refer readers to~\cite{Atefeh-ci15,hasan2018survey,saeed2019s}.

\subsection{Location-specific Event Detection}

Various approaches have also been developed for event detection in the spatial, textual (i.e. semantic) and temporal context, with many of these considering these different aspects separately \cite{zhang2016geoburst,zhang2018geoburst}. However, there are only limited works which combine spatio-temporal information for event detection. In \cite{huang2018spatial}, spatio-temporal events are detected by clustering the geotagged tweets, followed by topic modelling using the summarized words in each estimated cluster. Similarly,~\cite{Kwan-IUI21,Kwan-ASONAM20} adopted an approach of identifying topics associated with specific locations by applying Latent Dirichlet allocation on tweets posted in the same locality. Others like~\cite{zhang2017Triovecevent} combine clustering techniques with embeddings of tweet location, time, and text for event detection. Although Twitter enables users to post tweets with their current locations (longitude and latitude), only an average rate of 0.85–3\% tweets being geotagged per day, around 7,000,000 geotagged tweets are posted per day \cite{li2017discover}. This characteristic where only a small proportion of the tweets are geotagged severely restricts the accuracy of spatial-based event detection approaches. 

Another key challenge in this research area is regarding the method by which the geographical area is partitioned for subsequent event detection. A typical approach is to utilize a uniform-grid approach, which applies an equal-width grid of a specific size over the data domain. However, this approach does not solve the problem for various reasons. First, a good method for choosing the grid size is required, which has not been adequately covered in the literature~\cite{6544872}. Second, fixed grid cells might not help in finding both local and global events. For example, using a low resolution grid for spatial data might capture only the global events occurring on the state or the country level, while a high resolution grid will detect the events on smaller scales (local events), i.e. within the community or the city where the grid cell ranges from 1km-50km. Another solution is to manually select a set of points of interests (POIs), where each POI is a fixed size grid cell. Following this approach, we can control the number of POIs based on tweets distribution density. For instance, areas in city centre might have many POIs with small grid cells, while areas far from the city might have few POIs with large grid size. But, having fixed POIs limits the location of detected events to the chosen POIs only. In addition, the manually selected POIs has to be done for each geographical area of analysis. 

\subsection{Differences with Earlier Works}

Our proposed method differs from these earlier works in the following ways. The existing works on general event detection aims to detect general events that are discussed on social media without identifying the locality of these events. In contrast, our proposed method aims to detect location-specific events, which are localized events happening within a specific area. Furthermore, the supervised approaches to general event detection requires implicitly labelled events that may not generalized well to unseen or new events, whereas our work do not require explicit event labels.

\section{Problem Statement}
\label{sectBgProblem}

In this section, we first introduce some basic notation and definitions used in our work, before formally defining the problem of spatio-temporal event detection.

\begin{table}[t]
	\centering
	
	\caption{High level notation}
	\small
	\label{tab:notations2}
	\setlength\tabcolsep{2pt}
	\setlength\extrarowheight{1pt}
	\begin{tabular}{cm{0.85\linewidth}}
		\hline\hline
		$p_x$ & A social media post $p_x$ \\ 
		$t_x$ & Timestamp of post $p_x$ \\
		$l_x$ & Location of post $p_x$ \\ 
		$f_x$ & Features of post $p_x$ \\ 
		$\mathcal{S}$ & A data stream of posts as a set, $\mathcal{S} = \{p_1, p_2, \dotsc, p_n\}$ \\ 
		$\mathcal{W}$ & A window (e.g. of size $m$), $\mathcal{W} = \{p_{n-m+1}, \dotsc, p_n\}\subset\mathcal{S}$ \\ 
		$\mathcal{E}$ & A set of posts $\mathcal{E} \subseteq \mathcal{W}$ that represents a spatio-temporal event \\ \hline\hline
	\end{tabular} 
	
\end{table}


\subsection{Preliminaries}

Table~\ref{tab:notations2} summarizes the key notations used in our work, which we elaborate next.

{\bf Definition 1 (Social Media Post)}: In our work, we aim to utilize geo-tagged social media as an input to our location-based event detection algorithm. As such, each social media post $p\in\mathcal{S}$ forms the basic component of our algorithm. We represent each social media post as $p = \langle t, l, f \rangle$, where each social media post $p$ is associated with a timestamp $t$, location $l$ and features $f$. The timestamp $t$ and location $l$ are straightforward representations of date/time and latitude/longitude coordinates but can be easily modified to other representations, e.g., unix timestamp and landmarks, instead of date/time and latitude/longitude coordinates. On the other hand, features $f$ can represent multiple aspects of different types of social media, e.g., text in a tweet, user tags for a photo, etc.

{\bf Definition 2 (Data Stream)}: Building upon Definition 1 (Social Media Post), we now have multiple social media posts arriving in a real-time data stream. Let $\mathcal{S} = \{p_1, p_2, \dotsc, p_n\}$ denote the first $n$ posts from the data stream, ordered temporally such that for $p_i$ and $p_j$ where $i<j$, $t_i \leq t_j$. Thus, data stream $\mathcal{S}$ represents a series of social media posts ordered in the sequence they arrived in.

{\bf Definition 3 (Current/Query Window)}: In the context of a Data Stream $\mathcal{S} = \{p_1, p_2, \dotsc, p_n\}$, we define a current/query window $\mathcal{W} = \{p_{n-m+1}, \dotsc, p_{n-1}, p_n\}$, where $\mathcal{W} \subseteq \mathcal{S}$. This current/query window represents the current set of social media posts from post $p_{n-m+1}$ to post $p_n$. For generalizability and flexibility, the window size can be based on either a fixed number of posts, $m> 0$, or a fixed duration between posts $p_{n-m+1}$ and post $p_n$, i.e. $t_n - t_{n-m+1}$. This current/query window allows a user to decide on the temporal resolution in which he/she wants to be able to detect an event. 

\subsection{Formal Problem Definition}

The focus of our work is to develop an algorithm for detecting spatio-temporal events from streaming social media, based on a provided set of current social media posts, i.e., the query/current window. We define a spatio-temporal event as a set of social media posts that represents an increase in activity across a period of time within the same locality, based on the current/query window.

Given a data stream of social media posts $\mathcal{S} = \{p_1, p_2, \dotsc, p_n\}$ and a query window $\mathcal{W} = \{p_{n-m+1},p_{n-m+2}, \dotsc, p_n\}$ that represents currently observed social media posts, we want to identify a set of posts $\mathcal{E} \in \mathcal{W}$ with the following goals:
\begin{itemize}
	\item Spatial Proximity, e.g. $\sum\limits_{p_x \in \mathcal{E}} \sum\limits_{p_y \in \mathcal{E}} dist(l_x, l_y)$ should be significantly smaller than that for the same number of posts drawn uniformly at random from $\mathcal{W}$.
	\item Temporal Proximity, e.g. $\sum\limits_{p_x,p_y \in \mathcal{E}} (t_y - t_x)$ (for consecutive $p_x$ and $p_y$), should be significantly smaller than that for the same number of posts drawn uniformly at random from $\mathcal{W}$.
	\item Significance, $\lvert \mathcal{E} \rvert$ should be as large as possible while maintaining Spatial and Temporal Proximity goals.
\end{itemize}
Here $dist(l_x, l_y)$ is the geographical distance between posts $p_x$ and $p_y$, while $(t_x - t_y)$ is the time difference between consecutive posts $p_x$ and $p_y$. In short, we are selecting a subset of social media posts that are representative of a spatio-temporal event, based on their spatial and temporal proximity.

\section{Proposed Algorithm}
\label{sectAlgorithm}

In our work, we address spatial proximity by considering windows that are defined in terms of a region, and we address temporal proximity by considering a sliding window and assessing the change in the number of posts in a given region for two consecutive windows. Formally, we consider a number of sliding windows, $\mathcal{W}_{i}(\Gamma_j)$, identified by an unbounded slide sequence number $i=1,2,\dotsc$ and finite set of regions $\Gamma_j$, $j=1,2,\dotsc,\gamma$, where $\bigcup \Gamma_j=\Gamma$. All of our sliding windows have a time duration of $T$ and slide increment $\Delta T$, with the head of the window being $T_i$, i.e. every sliding window $i$ covers time interval $[T_i-T,T_i)$. In this way, we define
\begin{equation}
\label{eq_w}
\mathcal{W}_{i}(\Gamma_j) = \mathcal{W}_{ij} = \left\{p_x \mid T_i-T \le t_x < T_i, l_x \in \Gamma_j, p_x \in \mathcal{S}\right\}. 
\end{equation}
Each region serves as a spatial proximity bound for the posts that it contains, in the sense that we can consider the posts being within a given region as satisfying the spatial proximity goal from the problem definition. There are many ways that regions can be selected, e.g. they could be a uniform mesh based partition of the space, or each region could be associated with a POI in the space (e.g. a region around a park or building), etc. In our work, we consider a multi-scale region selection approach based on a quad-tree division of space; in this case regions are overlapping with some regions subsuming others. In previous work, we have also considered the POI region selection approach and we make comparisons between them in this paper.

\begin{figure}[!th]
	\centering
	\includegraphics[width=.9\linewidth]{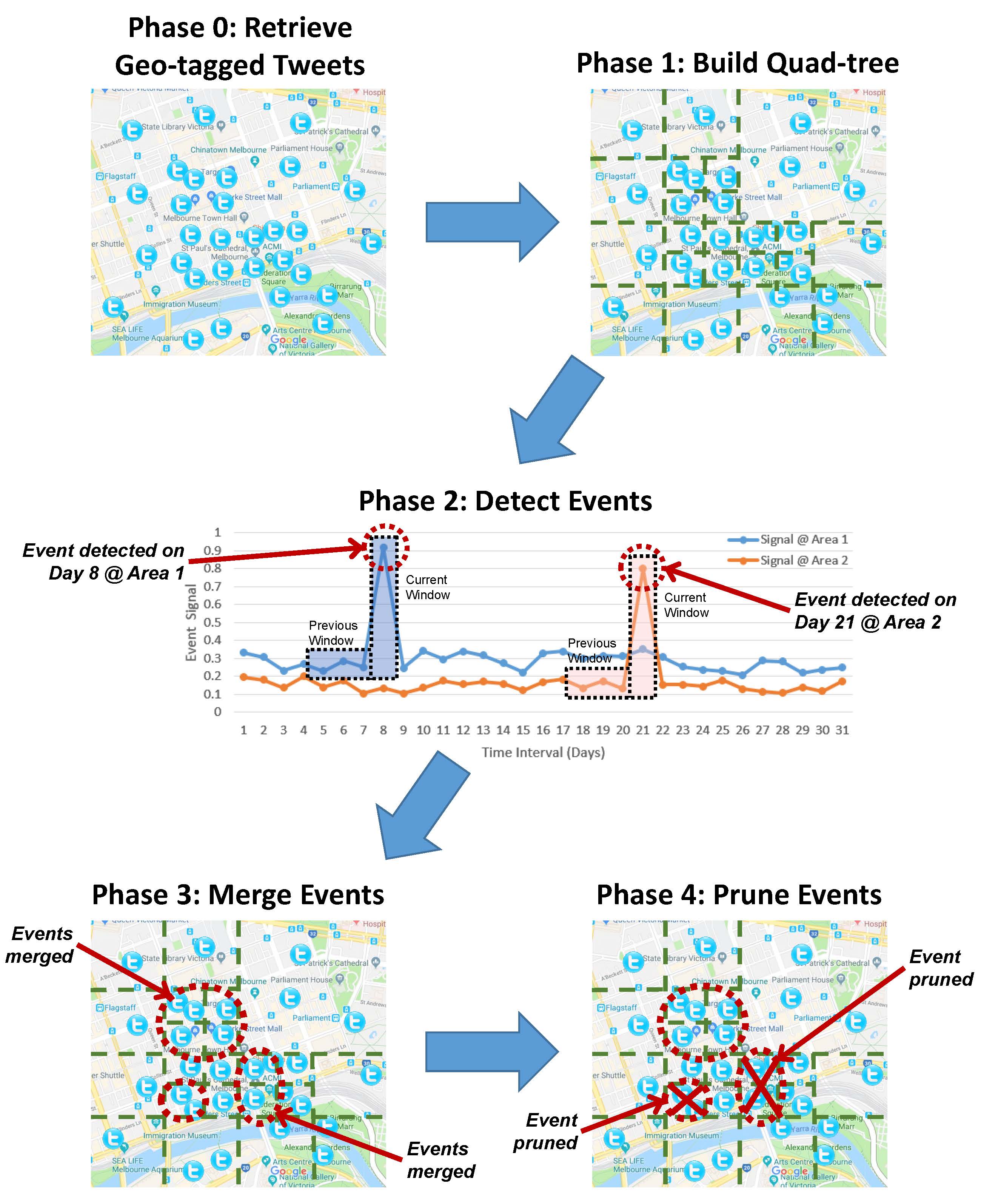}
	\caption{Overview of our proposed Spatio-temporal Online Event Detection Algorithm.}
	\label{fig:framework}
\end{figure}

In order to assess the change in the number of posts from one window to the next, we assume that the number of posts arriving in a given time interval has a Poisson distribution, and we thereby assign an estimate $\Delta T$-arrival rate of posts for each region based on its sliding window:
\begin{equation}
\label{lambda_equation}
\lambda_{ij} = \frac{\lvert \mathcal{W}_{ij} \rvert\Delta T}{T},
\end{equation}
where $|x|$ is the cardinality of set $x$. Finally, as the basis for event detection in each region, for each window $\mathcal{W}_{ij}$, we consider the observed number of posts in the slide increment interval $[T_i,T_i+\Delta T)$,
\[
C_{ij}=\Big\lvert\big\{ p_x \mid T_i \le t_x < T_i+\Delta T, l_x \in \Gamma_j, p_x \in \mathcal{S}\big\}\Big\rvert,
\]
and make use of the Poisson p.m.f.:
\begin{equation}
\label{eq_poisson}
\mathbb{P}\big[C_{ij};\lambda_{ij}\big] = P_{ij} = e^{\lambda_{ij}} \frac{\lambda^{C_{ij}}}{C_{ij}!}.
\end{equation}
If $P_{ij}$ is significantly low (below a threshold) then we consider the possibility that region $j$ has exhibited an event and we consider the posts within the slide increment to potentially be comprising that event. The details of our approach, called Spatio-temporal Online Event Detection Algorithm (Algorithm~\ref{main_algorithm}), includes more aspects that are explained next, such as: \emph{(1)} building a multiscale spatial resolution grid using the quad-tree method, \emph{(2)} event detection using the Poisson model and signal smoothing, \emph{(3)} event merging and \emph{(4)} event pruning. Generally, our algorithm maintains an unbounded set of detected events $\mathcal{E}$ found in the unbounded stream $\mathcal{S}$. Figure~\ref{fig:framework} provides an overview of our algorithm and the detailed explanation for each phase is provided in the following subsections. As well, Table~\ref{tab:algonotation} provides an overview of the notation used in the algorithm.

\begin{table}[t]
	\centering
	
	\caption{Additional notation}
	\small
	
	\label{tab:algonotation}
	\setlength\tabcolsep{2pt}
	\setlength\extrarowheight{1pt}
	\begin{tabular}{cm{0.85\linewidth}}
		\hline\hline
		$\Gamma_j$ & region $j$ where $\Gamma=\bigcup\Gamma_j$ contains all posts in $\mathcal{S}$ \\ 
		$\mathcal{W}_{ij}$ & the set of posts in the sliding window at interval $i$ for region $j$ \\
		$C_{ij}$ & number of posts in the window slide increment interval $i$ for region $j$ \\ 
		$\lambda_{ij}$ & estimate rate of posts at interval $i$ for region $j$ \\ 
		$P_{ij}$ & Poisson signal at interval $i$ for region $j$ \\
		$\tau_1$ & Poisson signal threshold \\ 
		$F_{ij}$ & event signal at interval $i$ for region $j$ \\
		$\tau_2$ & event detection threshold\\ 
		$\alpha$ & event signal decay parameter \\ 
		${\theta}_{duration}$ & event duration threshold \\ 
		${\theta}_{entity}$ & minimum entities threshold \\ 
		${\theta}_{area}$ & quad-tree node region area threshold \\ 
		${\theta}_{count}$ & quad-tree node post count threshold \\ \hline\hline
	\end{tabular} 
	
\end{table}

\begin{algorithm}
	\scriptsize
	\caption{Spatio-temporal Online Event Detection Algorithm} \label{main_algorithm}
	
	\begin{algorithmic}[1]
		\BState \textbf{Global constants (inputs):} $\Gamma$,$T$,$\Delta T$,${\theta}_{area}$,${\theta}_{count}$,$\tau_1$,$\tau_2$,$\alpha$,${\theta}_{duration}$,${\theta}_{entity}$
		\BState \emph{\textbf{Output}: $\mathcal{E}$ }	 
		\BState \emph{\textbf{Initialise}}: $i \gets 1,  \mathcal{E} \gets [] $
		\BState \emph{\textbf{Repeat}}:
		\Indent
		\State $qTree$ $\gets$ \textproc{quadTree}($i$,$\Gamma_1=\Gamma$) \Comment{Phase 1: Build quad-tree}
		\State $\mathcal{E}_i \gets$ \textproc{eventDetection}($i$,$qTree$) \Comment{Phase 2: Detect events}
		\State $\mathcal{E} \gets$ \textproc{eventMerge}($\mathcal{E}$,$\mathcal{E}_i$) \Comment{Phase 3: Merge events}
		\State $\mathcal{E} \gets$ \textproc{eventPrune}($\mathcal{E}$) \Comment{Phase 4: Prune events}
		\State $i \gets i+1$

		\EndIndent

		\Algphase{Phase 1: Build Quad-tree}
		
		\Function{quadTree}{$i$,$\Gamma_{x}$}
		\State $N\gets\{(\lambda_{ix},C_{ix})\}$ \Comment{This node is synonomous with region $x$}
		\If {$\left\lvert\mathcal{W}_{ix}\right\rvert > \theta_{count}$ and area of $\Gamma_x\geq \theta_{area}$}
		\State Subdivide $\Gamma_x$ into $\Big\{\Gamma_{x1},\Gamma_{x2},\Gamma_{x3},\Gamma_{x4}\Big\}$ 
		\State $N$ $\gets$ $N$ $\cup$ \textproc{quadTree}($i$,$\Gamma_{x1}$)
		\State $N$ $\gets$ $N$ $\cup$ \textproc{quadTree}($i$,$\Gamma_{x2}$)
		\State $N$ $\gets$ $N$ $\cup$ \textproc{quadTree}($i$,$\Gamma_{x3}$)
		\State $N$ $\gets$ $N$ $\cup$ \textproc{quadTree}($i$,$\Gamma_{x4}$)
		\EndIf
		\State \Return N
		\EndFunction
		
		\Algphase{Phase 2: Event Detection}
		\Function{eventDetection}{$i$,$qTree$}
		\State $\mathcal{E}\gets\{\}$
		\For{node $(C_{ix},\lambda_{ix})$ in $qTree$}
		\State Compute $P_{ix}$ using Eq.~\ref{eq_poisson}
		\State Compute $F_{ix}$ using Eq.~\ref{eq_smoothing}
		\If{$ F_{ix} \ge \tau_2$}
		\State $e \gets (\Gamma_{ix}, T_i,T_i+\Delta T, \Delta T,\mathcal{W}_{ix},P_{ix})$ 
		\State \Comment{region, start, end, period, posts, signal}
		\State $\mathcal{E}\gets\mathcal{E}\cup\{e\}$
		\EndIf
		\EndFor
		\State \Return $\mathcal{E}$
		\EndFunction

		\Algphase{Phase 3: Merge Events}
		
		\Function{eventMerge}{$\mathcal{E}$, $\mathcal{E}_i$}
		\For{event $e=(rg,st,en,pe,po,si) \in \mathcal{E}_i$}
		\If {$ e^\prime=(rg^\prime,st^\prime,en^\prime,pe^\prime,po^\prime,si^\prime) \in \mathcal{E}$ where $rg^\prime=rg$ and $en^\prime=st$}
		\State $e^\prime\gets \left(rg^\prime,st^\prime,en^\prime+\Delta T,pe^\prime+\Delta T,po\cup po^\prime,\frac{si+si^\prime}{2}\right)$ 
		\State \Comment{Updates the existing event in $\mathcal{E}$}
		\Else
		\State $\mathcal{E} \gets \mathcal{E} \cup \{e\} $
		\EndIf
		\EndFor
		\State \Return $\mathcal{E}$
		\EndFunction
		
		\Algphase{Phase 4: Prune Events}
		\Function{eventPrune}{$\mathcal{E}$}
		\For{event $e=(rg,st,en,pe,po,si) \in \mathcal{E}$}
		\If{$pe< \theta_{duration}$}
		\State $\mathcal{E}\gets\mathcal{E}/\{e\}$ \Comment{The duration of the event is insufficient}
		\State Continue
		\EndIf
		\If{$\exists e^\prime=(rg^\prime,st^\prime,en^\prime,pe^\prime,po^\prime,si^\prime) \in \mathcal{E} \mid si^\prime < si, rg^\prime \cap rg \neq \emptyset$}
		\State $\mathcal{E}\gets\mathcal{E}/\{e\}$ \Comment{The event is subsumed by a stronger event}
		\State Continue
		\EndIf
		\If{$|uniqueEntities(e)| < \theta_{entity}$}
		\State $\mathcal{E}\gets\mathcal{E}/\{e\}$ \Comment{The event is not semantically consistent}
		\State Continue 
		\EndIf
		\EndFor
		\State \Return $\mathcal{E}$
		\EndFunction
		
	\end{algorithmic}
\end{algorithm}

\subsection{Phase 1: Build Quad-tree}
In this phase, we use the quad-tree method for spatial decomposition~\cite{finkel1974quad,wang2003quadtree}. This method has been used in a variety of applications including image processing, computer graphics, geographic information systems and robotics~\cite{samet1984quadtree,rosenberg1985geographical}. We construct a quad-tree at each time interval $i$. The quad-tree in two dimensional space starts with a large rectangular region, in our work $\Gamma_1=\Gamma$, which represents the root of the quad-tree. The root region $\Gamma_1$ is subdivided into four equal sized regions $\{\Gamma_{11},\Gamma_{12},\Gamma_{13},\Gamma_{14}\}$ and each subregion is recursively subdivided, i.e. creating $\{\Gamma_{111},\Gamma_{112},\dotsc\}$, and so on. Subdivision of a region $x$ only occurs if both $|\mathcal{W}_{ix}|>\theta_{count}$ posts and the area of region $x$ is at least $\theta_{area}$. These constraints limit the minimum spatial resolution. As the quad-tree is constructed we also compute $\lambda_{ij}$ and $C_{ij}$ for each node, including internal nodes; here node is synonymous with region in that region $j$ is node $j$.

\subsection{Phase 2: Event Detection}

For a sliding window interval $i$ and all regions $j$ (including those at internal nodes of the quad-tree), we use the Poisson distribution~\cite{yamane1973statistics,sokal1969principles,patel1976handbook} to measure how likely the observed number of posts, $C_{ij}$, is for the slide increment $\Delta T$ that immediately follows the sliding window. The estimate arrival rate of posts is computed as in Eq.~\ref{lambda_equation} and the probability, $P_{ij}$, of observing $C_{ij}$ posts in time $\Delta T$ is computed as in Eq.~\ref{eq_poisson}. The more unlikely the observation, which may result from a significantly large increase or decrease in posts from the mean, the more we consider the posts (or lack thereof) to comprise an event. Therefore regions with $P_{ij}<\tau_1$, a constant threshold, could be flagged as potential regions for events. To compensate for sparse and/or incomplete data, where the stream of posts may not have a significantly strong representation of social media posts, we ``smooth" the Poisson signal by computing an exponential decaying average \emph{event signal}, $F_{ij}$:
\begin{equation}
\label{eq_smoothing}
\begin{split}
\delta_{ij}=
\begin{cases}
\frac{\tau_1-P_{ij}}{\tau_1}, & \text{if}\ P_{ij} < \tau_1 \\
0, & \text{otherwise}
\end{cases}  
\\ \newline 
F_{ij} = \alpha F_{i-1,j} +  (1- \alpha)\delta_{ij}
\end{split}
\end{equation}
where $\delta_{ij}$ is the scaled Poisson signal, $F_{ij}$ and $F_{i-1,j}$ are the event signal values for node (region) $j$ at the $i$ and $i-1$ intervals respectively, and $0\leq \alpha\leq 1$ is a constant decay parameter. Finally, if $F_{ij}\geq \tau_2$, an event detection threshold, we flag the posts, or more specifically the interval and region, as comprising an event.

\subsection{Phase 3: Merge Events}

Each event found in the previous Event Detection phase has a different spatial resolution and a fixed temporal scale ($\Delta T$). In this phase, we construct events with multiscale temporal resolution using a merging method. Events at the same region that occur at consecutive time intervals are merged. This gives an estimate for the period of time during in which an event is highlighted rather than assuming a predefined fixed duration ($\Delta T$). For instance, if two events $e_1$ and $e_2$ occur in the same region at time intervals $i$ and $i+1$ in order, then both events are combined to one event with period of $2 \Delta T$. When merging we combine the posts and average the signal strength for the merged event. 

\subsection{Phase 4: Prune Events}

To further increase the precision of our event detection and to handle spatio-temporal events that occur over a changing resolution we prune events after merging them.
First, we only select events with duration $\ge \theta_{duration}$ to be included in the final set of detected events. The idea is that the longer the event duration, the more reliable and accurate it is. In other words, regions/nodes which are flagged for short periods are most likely to be noise (i.e. false positives). Second, the fact that we compute the signal for all quad-tree nodes (i.e. both internal and leaf nodes), leads to the propagation of some flagged events over the different tree levels (i.e. multi spatial resolution). So if an event is detected at the same time on different tree levels, we only keep the node with the strongest signal. In other words, if overlapping tree nodes (i.e. parent, child, grand child, and so on) are flagged as events from time $t_1$ to $t_2$, then we select the node (i.e. region) with the strongest signal to be the spatial resolution of the final detected event. This gives us a set of unique events which happened at different spatial and temporal resolutions. Third, we utilize the entities in the social media posts to detect and eliminate spam or fake events. We extract the set of unique entities (which may be keywords, mentions, hashtags, etc., depending on the type of social media post) across the posts in the event. If the size of the set is less than $\theta_{entity}$ then we remove the event.

\section{Experimental Design}
\label{sectExpDesign}
In this section, we describe our datasets and give an overview of the evaluation metrics and baseline algorithm used in our experimental methodology.

\subsection{Dataset and Data Collection}
\label{sec:dataset}

To demonstrate the generalizability of our proposed algorithm, we perform our experimental evaluation on two datasets based on Twitter and Flickr. 

For our Twitter dataset, we performed a two-stage collection of tweets, similar to~\cite{lim-smartcity19}. We first used the Twitter REST API to retrieve all geo-tagged tweets posted by users in Melbourne in 2017. As we focus on geo-tagged tweets, this collection process resulted in $203,519$ geotagged tweets by $22,264$ different users.

For our Flickr dataset, we focused on geo-tagged photos posted in four large cities, namely Melbourne, London, New York and Paris. These geo-tagged photos were extracted from the Yahoo! Flickr 100M Creative Commons (YFCC100M) dataset~\cite{yfcc100m,thomee-cacm16}. The YFCC100M dataset comprises 100M geo-tagged photos and videos along with their meta-data such as latitude/longitude coordinates, date/time taken, photo name, user description, assigned tags, etc.

\subsection{Evaluation Methodology}

In our experiments, we evaluate the various algorithms using the standard metrics of precision and recall, and our proposed metric of strength index. Precision and recall are defined based on the common definitions of true and false positives and true and false negatives, commonly used in confusion matrix. In the context of our study, they are defined as:
\begin{itemize}
    \item True Positive ($tp$): Detecting an event when the event has occurred in real-life.
    \item False Positive ($fp$): Detecting an event when the event did not occur in real-life.
    \item True Negative ($tn$): Not detecting an event when the event did not occur in real-life.
    \item False Negative ($fn$): Not detecting an event when the event has occurred in real-life.
\end{itemize}

\subsubsection{\textbf{Precision}}
We use precision to measure the ratio of correctly detected events (true positives) to the total detected events, by using the formula in Eq. \ref{eq:precision}. 

\begin{equation}
\label{eq:precision}
{\text{Precision}}={\frac {tp}{tp+fp}}
\end{equation}

where $tp$ and $fp$ are true and false positives respectively. 

The absence of ground truth labels makes the task of computing precision very challenging. As it is impractical to manually label the overly large number of events in the dataset, we propose a semi-automated assessment methodology using Google search results where each event is assigned 1 if it is a true event, 0 otherwise. To do so, we first query Google using the top k entities as well as the date-time of each detected event. We use Google query results to decide whether an event is True or False, with 1 and 0 referring to true and false event, respectively. If we do not get any useful information about the event from Google, then we manually look at the posts of the event to decide if it is a personal/private event, spam or wrong event. 

\subsubsection{\textbf{Recall}}
Recall is also calculated using Eq. \ref{eq:recall}, which reflects the ability of the model to find all actual events within a dataset. In the context of event detection, recall measures the percentage of detected events with respect to important events/news appearing on a real-world news headlines.  
\begin{equation}
\label{eq:recall}
{\text{Recall}}={\frac {tp}{tp+fn}}
\end{equation}

where $tp$ and $fn$ are true positives and false negatives respectively.  

Similar to precision, we perform a manual assessment for recall due to the absence of ground truth events. This is done by using Google search engine to select the most common events appearing on the news headlines for the days corresponding to the analysis. This includes festivals, public holiday events and international performances occurring in the area of analysis. 
Each event is represented by a list of entities, which are used to manually decide whether an event is detected by our method or not. 

\subsubsection{\textbf{Strength Index (SI)}}
To examine if the posts assigned to an event $e=(region$, $start$, $end$, $period$, $posts$, $signal)$ are relevant or not, we introduce a metric, which we refer to as the event strength index ($SI$). $SI$ is the fraction of the retrieved top entities to the total count of event posts. We use $SI$ as an indicator of how important/precise a reported event is. For an event $e$ with total number of posts $C=\lvert posts \rvert$ and ${\chi}_i$ being the $i$-th most frequent entity (could be hashtags and mentions for twitter or image tags and description for Flickr), we calculate $SI$ using the following formula:
\begin{equation}
\label{eq:strength}
{\text{Strength Index (SI)}}={\frac {\sum_{i=1}^{k}{C_{{\chi}_i}}}{C}},
\end{equation}
for constant $k>0$, where $C_{{\chi}_i}$ is the number of posts that contain ${\chi}_i$. $SI$ ranges from 0 to $k$, where $k$ is the number of top entities. We obtain a small value for $SI$ ($\ll 1$), when the top entities do not match the context of the detected event or if they are relevant but with a small number of occurrences. For example, a value of 0 for $SI$ means that all posts for an event $e$ are irrelevant, while a value of $k$ means that all event posts contain at least one occurrence of each top entity.

\subsection{Baseline Algorithm}

To show the effectiveness of the proposed method, we compare it with two baseline  event detection algorithms. The first baseline is based on POIs and the second one uses clustering approach for real-time event detection. More details about each baseline are provided in the following sections.

\begin{table}
	\centering
	\caption{Parameters used in the proposed method}
	\small
	\label{tab:parameters}
	\setlength\tabcolsep{5pt}
	\setlength\extrarowheight{1pt}
	\begin{tabular}{ll}
		\hline\hline
		\textbf{\textit{Method}} &\textbf{\textit{Parameter = Value}} \\
		\hline\hline
		\multirow{3}{*}{Quad-tree}  & ${\theta}_{count}$= $20$ \\ 
		& ${\theta}_{area}$ = 0.001 sqkm \\ \hline
		\multirow{5}{*}{Poisson model} & T  = 3 days \\ 
		& $\Delta T$ = 10 minutes \\ 
		& signal threshold $\tau_1$  =$ 0.01$ \\ 
		& event detection threshold $\tau_2$  =$ 0.4$ \\ 
		& event signal decay coefficient $\alpha$ = $0.5 $ \\ \hline 
		\multirow{2}{*}{Event filtering} & duration threshold ${\theta}_{duration}$= $50$ minutes \\ 
		& minimum top entities ${\theta}_{entity}$ = $2$ \\ \hline\hline	
	\end{tabular} 
\end{table}

\subsubsection{Location-based Event Detection using Points of Interests (POIs)}

POIs have been frequently used in location-based recommendation~\cite{Cai-esa18,chen-cikm16,brilhante-ipm15} and various types of event detection techniques based on POIs~\cite{lin2011event}.
Building upon these works, we develop a baseline using the similar idea of tagging geo-tagged social media to POIs. Similar to these earlier works, we obtain a list of known and popular POIs for each city from their respective Wikipedia entries. In this baseline algorithm, we utilize a spatial representation of tweets based on their proximity ($<$100m) to known POIs, instead of assigning tweets to dynamically-sized grids based on quad trees. The remaining steps of computing Poisson signals and determining event duration remain the same as previously described in Section~\ref{sectAlgorithm}. 

\begin{table}[!t]
	\centering
	\caption{Parameter selection: Poisson signal threshold ($\tau_1$)}
	\small
	\label{tab:tau_1}
	\setlength\tabcolsep{3pt}
	\setlength\extrarowheight{2pt}
	\begin{tabular}{p{4.2cm}ccccccc}
		\hline\hline
		
		\textbf{\textit{	$\tau_1$ value}} & \textbf{\textit{0.01}} &\textbf{\textit{ 0.005}} & \textbf{\textit{0.0025}} & \textbf{\textit{0.0013}} & \textbf{\textit{0.0006}} & \textbf{\textit{0.0003}} &\textbf{\textit{ 0.0001}}\\ \hline \hline
		\textbf{\textit{Phase}} & \multicolumn{7}{c}{Poisson signal } \\ \hline 
		Signal thresholding & 6846 & 4821 & 3546 & 2458 & 1884 & 1557 & 1288\\ 
		Merging adjacent $\Delta T$ events & 6144 & 3882 & 2748 & 1763 & 1268 & 992 & 761\\ 
		Event duration D $ \ge \theta_{duration}$ & 40 & 27 & 16 & 11 & 7 & 3 & 2\\ 
		Filtering  propagated events & 21 & 17 & 11 & 8 & 5 & 1 & 1\\ 
		Spam filtering & 16 & 11 & 8 & 6 & 5 & 1 & 1\\ \hline 
		\textbf{\textit{Phase}} & \multicolumn{7}{c}{Smoothing signal}\\ \hline 
		Signal thresholding & 3332 & 2462 & 1931 & 1573 & 1315 & 1060 & 927\\ 
		Merging adjacent $\Delta T$ events & 2292 & 1604 & 1188 & 923 & 726 & 519 & 424\\ 
		Event duration D $ \ge \theta_{duration}$ & 51 & 29 & 16 & 8 & 6 & 3 & 2\\ 
		Filtering  propagated events & 29 & 22 & 11 & 6 & 5 & 2 & 1\\ 
		Spam filtering & 23 & 15 & 8 & 4 & 4 & 2 & 1\\ \hline \hline
		
	\end{tabular}
\end{table}

\subsubsection{Incremental Clustering for Real-time Event Detection}

Among the existing event detection techniques and algorithms discussed in the introduction, we select the clustering based event detection approach proposed in \cite{Andrienko-2015} for comparison with our approach. The reason is that it is very closely related to our introduced problem. The approach detects significant clusters that are sufficiently dense and large in streams of spatial events, with the advantage of tracking cluster evolution over time. 

Given a list of active data points (i.e. spatial events) that occur in the interval $[t_c - \Delta T, t_c]$, where $t_c$ is the current time and $\Delta T$ is a maximal temporal gap, the algorithm finds the set of significant clusters by repeatedly extracting a set of event circles and unions every time tick $t$. In which, an event circle C is a group of active events that fits in a circle with maximal radius R. While a union is a set of event circles that have at least $K$-overlapped events (i.e. $K$-connecting events). Following this, the algorithm finds all significant clusters, where a significant event cluster is a union that includes at least N spatial events, i.e. minimal cluster size. The values for parameters $t$, $\Delta T$, $R$, $K$, and $N$ are user-specified. More details about the algorithm can be found in \cite{Andrienko-2015}.

\section{Results}
\label{sectExperiments}
In this section, we evaluate the proposed method in four different aspects. Firstly, we present a preliminary analysis of the proposed method (Section \ref{sec:res-proposedMethod}). Secondly, we present a detailed comparative analysis with the baseline algorithm (Section \ref{sec:res-compResults}). Thirdly, using the tweets over a period of one-year we evaluate our algorithm based on the precision, recall and strength index as statistical metrics (\ref{sec:res-caseStudyTwitter}). Finally, we show a case study of event detection using Flickr image dataset (Section \ref{sec:res-caseStudyFlickr}). 

\subsection{Preliminary Analysis}
\label{sec:res-proposedMethod}
In this section, we use a subset of the collected tweets to evaluate the individual phases of the proposed method. We extract January-2017 Melbourne tweets which contains 23,327 geotagged tweets for 5,427 users. First, quad-tree is used to construct multiscale spatial grid. Then, events are detected using Poisson model. Following this, a smoothing function is applied for accurate estimation of event duration. Finally, a false positives removal phase is performed to eliminate both falsely highlighted events and spam events. Table~\ref{tab:parameters} presents the different parameters used in the proposed method. The parameters are chosen after several experiments, to achieve the best performance. Figure~\ref{fig:events-15jan} shows an example of the quad-tree results. 

The Poisson signal threshold $\tau_1$ has a strong impact on the results of our proposed method, and so we explored this parameter and show in Table~\ref{tab:tau_1} the number of events detected for variable values for $\tau_1$ at different phases. Combined with Table~\ref{tab:tau_1_just}, it is clear that reducing $\tau_1$ reduces the total number of detected events while keeping the strongest events. For our case studies, we chose $\tau_1 = 0.01$ so we can get reliable estimation of events in terms of both precision and recall.

\begin{table}[!t]
	\centering
	\caption{Sample detected events for different Poisson signal thresholds}
	\small
	\label{tab:tau_1_just}
	\setlength\tabcolsep{7pt}
	\setlength\extrarowheight{2pt}
	\begin{tabular}{lccp{6cm}}
		\hline\hline
		
		Event time & C & Area & Top hashtag/mentions \\ \hline \hline
		
		\multicolumn{4}{c}{Events using smoothing method with $\tau_1 = 0.00125$ } \\ \hline 
		
		-S 01-18 10:10 -D 50 & 27 & 6.969 & \#funk, @choiproductions, \#nofilterwtf, @jeffreymayfield, \#saxophone \\ 
		-S 01-21 07:50 -D 70 & 20 & 0.435 & \#gostars, @mcg, \#mcg, \#bbl06, \#melbourne \\ 
		-S 01-22 02:10 -D 70 & 18 & 0.109 & \#50th   \\ 
		-S 01-22 08:20 -D 60 & 15 & 0.109 &    \\ 
		-S 01-29 12:10 -D 60 & 72 & 1.741 & \#ausopen, \#australianopen, \#federer, \#rogerfederer, \#champion \\ 
		-S 01-29 12:10 -D 70 & 73 & 0.109 & \#ausopen, \#australianopen, \#federer, \#champion, \#rogerfederer \\ \hline

		\multicolumn{4}{c}{ Events using smoothing method with $\tau_1 = 0.000625$ } \\ \hline 
		-S 01-21 08:10 -D 50 & 14 & 0.435 & \#gostars, \#mcg, @mcg, \#bbl06, \#melbourne \\ 
		-S 01-29 12:10 -D 70 & 73 & 0.109 & \#ausopen, \#australianopen, \#federer, \#champion, \#rogerfederer \\ \hline 
		
		\multicolumn{4}{c}{Events using Poisson signal with $\tau_1 = 0.0003125$} \\ \hline 
		
		-S 01-29 12:10 -D 50 & 61 & 0.109 & \#ausopen, \#australianopen, \#federer, \#rogerfederer, \#18 \\ \hline \hline
	\end{tabular}
\end{table}

\begin{figure*}[!t]
	\centering
	\setlength\tabcolsep{2pt}
	\begin{tabular}{ccc}
		\includegraphics[trim={2cm 0cm 0cm 0cm},clip,width=.33\linewidth]{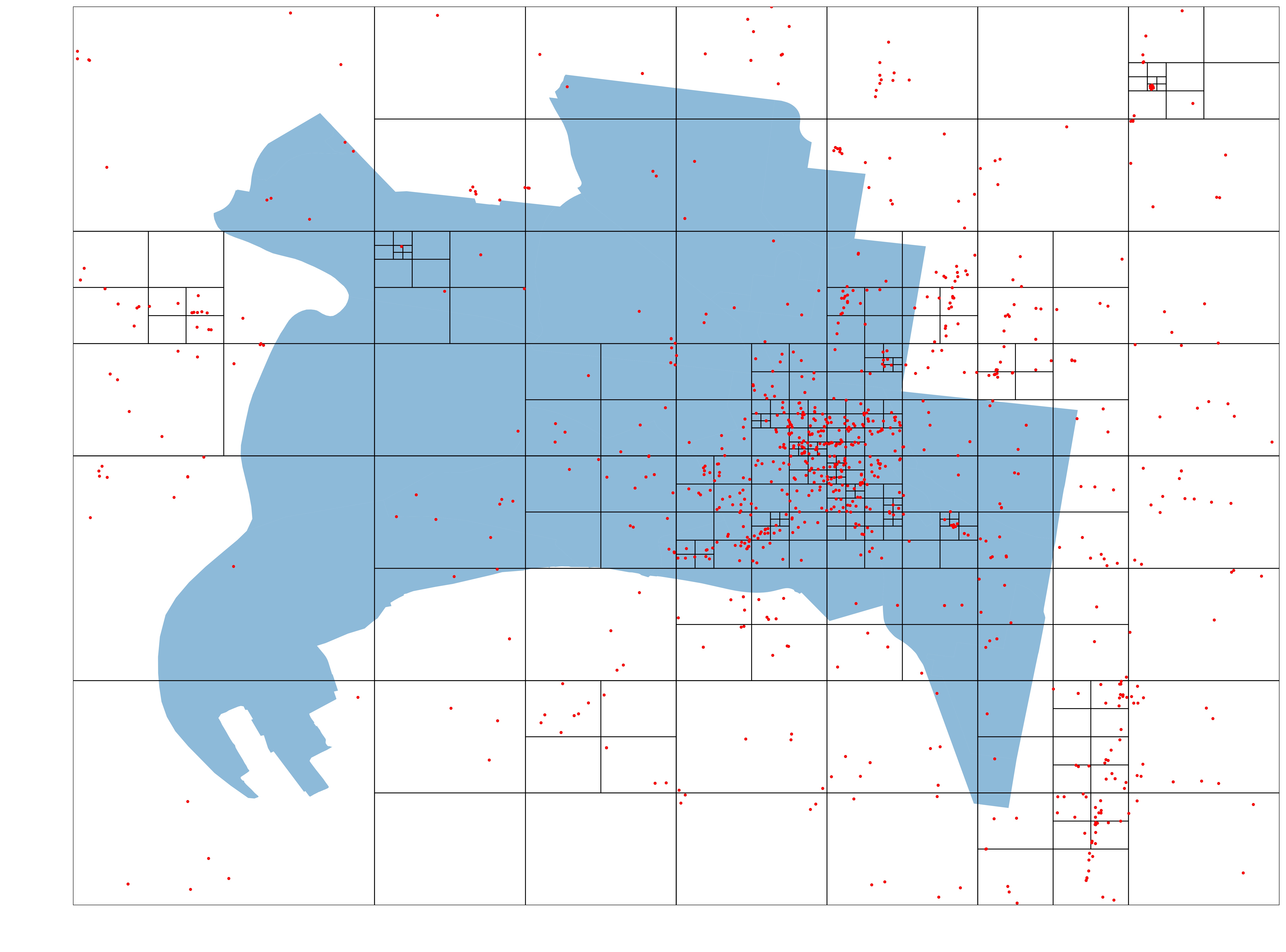} &
		\includegraphics[trim={10cm 5cm 10cm 9cm},clip,width=.33\linewidth]{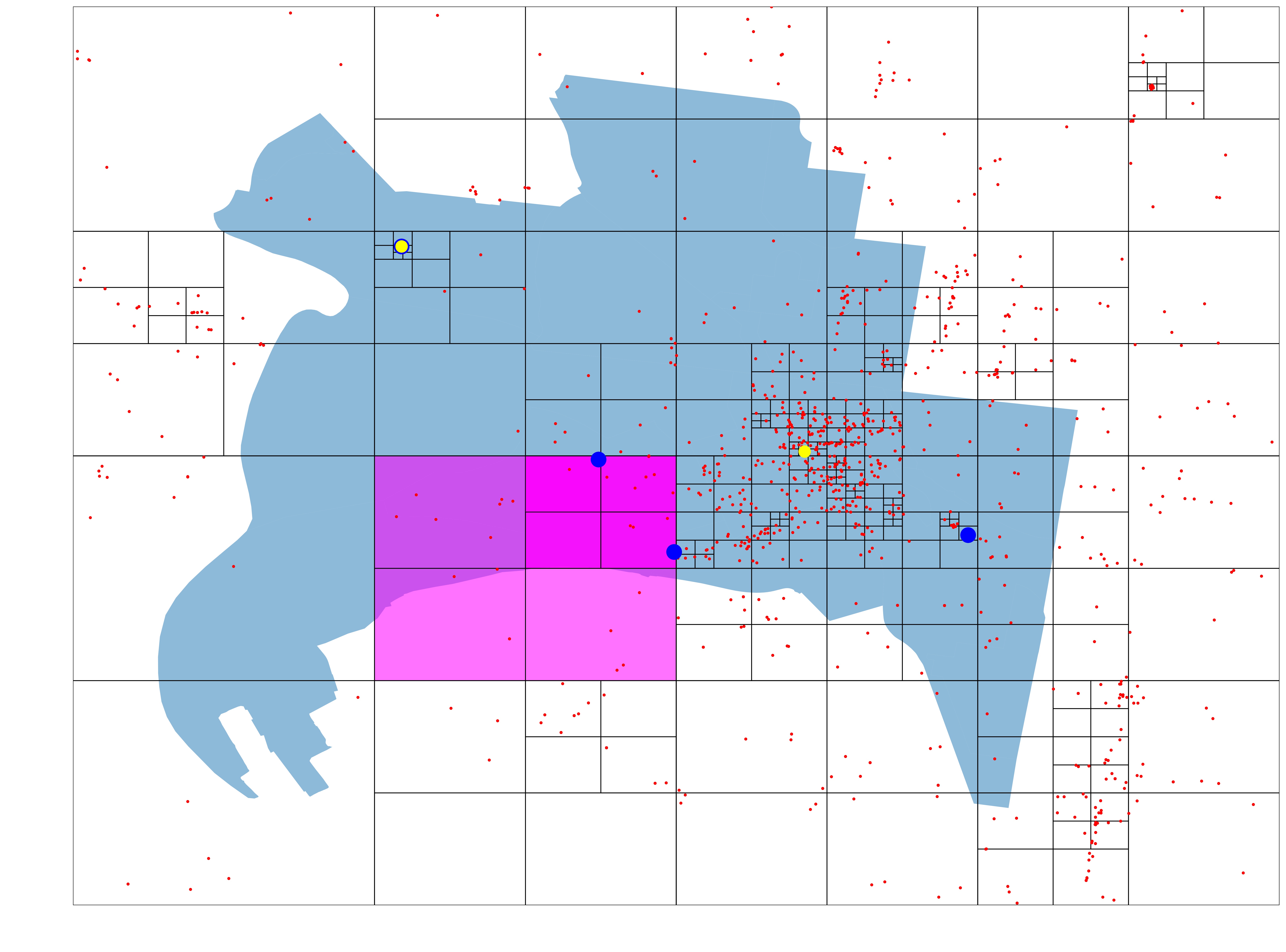} &	
		
		\includegraphics[trim={0cm 0cm 0cm 0cm},clip,width=.3\linewidth]{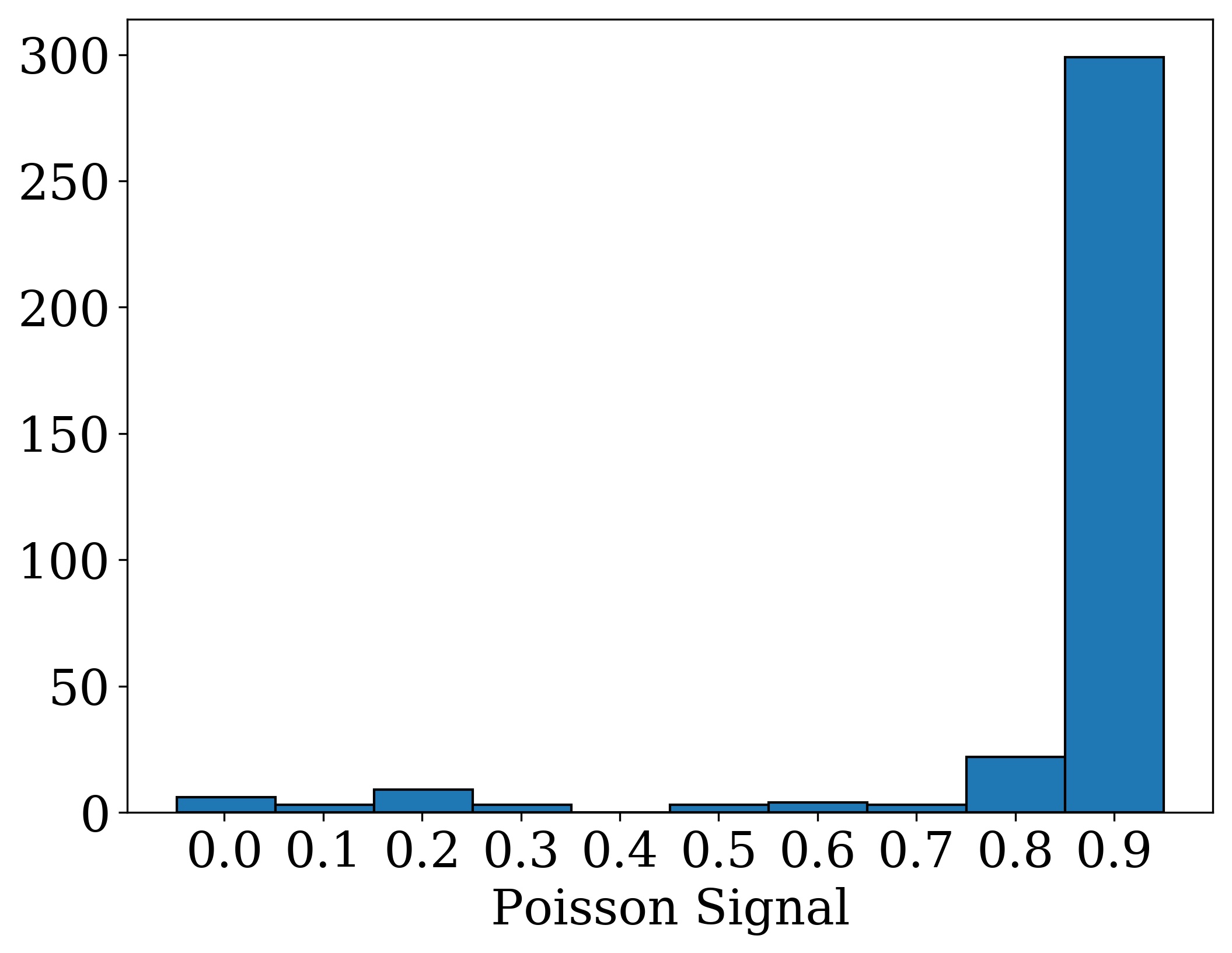} \\
		\text{(a)} & \text{(b)} & \text{(c)} \\

	\end{tabular}
	\caption{Event detection results at time-stamp $t = 2017-01-15$ $7pm$: (a) Constructed quad-tree using tweets in the interval [$t-T:t$) where $T=3$-days,  (b) Flagged events with Poisson signal, and (c) Distribution of Poisson signals for all  nodes.}	
	\label{fig:events-15jan}

\end{figure*}	

\begin{table}[!th]
	\centering
	\caption{Event detection results for Flickr and Twitter}
	\small
	\label{tab:flickr_twitter_details}
	\setlength\tabcolsep{4pt}
	\setlength\extrarowheight{2pt}
	\begin{tabular}{p{.35\linewidth}ccccc}
		\hline\hline
		
		\textbf{\textit{Case Study}} & \textbf{\textit{Twitter}}  &\multicolumn{4}{c}{ \textbf{\textit{Flickr}} } \\ 
		
		\textbf{City} &	Melbourne & Melbourne	&London	&New York&	Paris \\ \hline
		
		Month-Year	&2017 &Jan-2013 &Jan-2012 &Jan-2013&	Jan-2013 \\ 
		Number of users&22264&	90&	500&	510&	200 \\ 
		Number of posts:& 203519	&995&	6691&	6180&	2778 \\ 
		
		Total 10-min intervals \& nodes	&15253260&230468&	1283452&	1294544&	641228 \\ 
		Total flagged nodes & 29520&726 &	4791	&4757	&1871 \\ 
		
		Merging adjacent nodes &25108&	165&	2706	&2641&	795 \\ 
		Duration filtering &299 &	19&	140&	145&	49 \\ 
		Filtering propagated events&158&	7	&67	&62	&22	 \\ 
		Filtering spam events &137 & 	 7 & 	67 &	56&	22 \\ \hline\hline
	\end{tabular} 
\end{table}

\subsection{Case Study: Twitter Dataset}

\label{sec:res-caseStudyTwitter}
\begin{figure}[!th]
	\centering
	\tiny
	\setlength\tabcolsep{.5pt}
	\setlength\extrarowheight{0pt}
	\begin{tabular}{p{0.12\linewidth}p{0.12\linewidth}p{0.12\linewidth}p{0.12\linewidth}p{0.12\linewidth}p{0.12\linewidth}p{0.12\linewidth}p{0.12\linewidth}}
		\multicolumn{2}{c}{
			\includegraphics[trim={15cm 12cm 8cm 15cm},clip,width=0.24\linewidth]{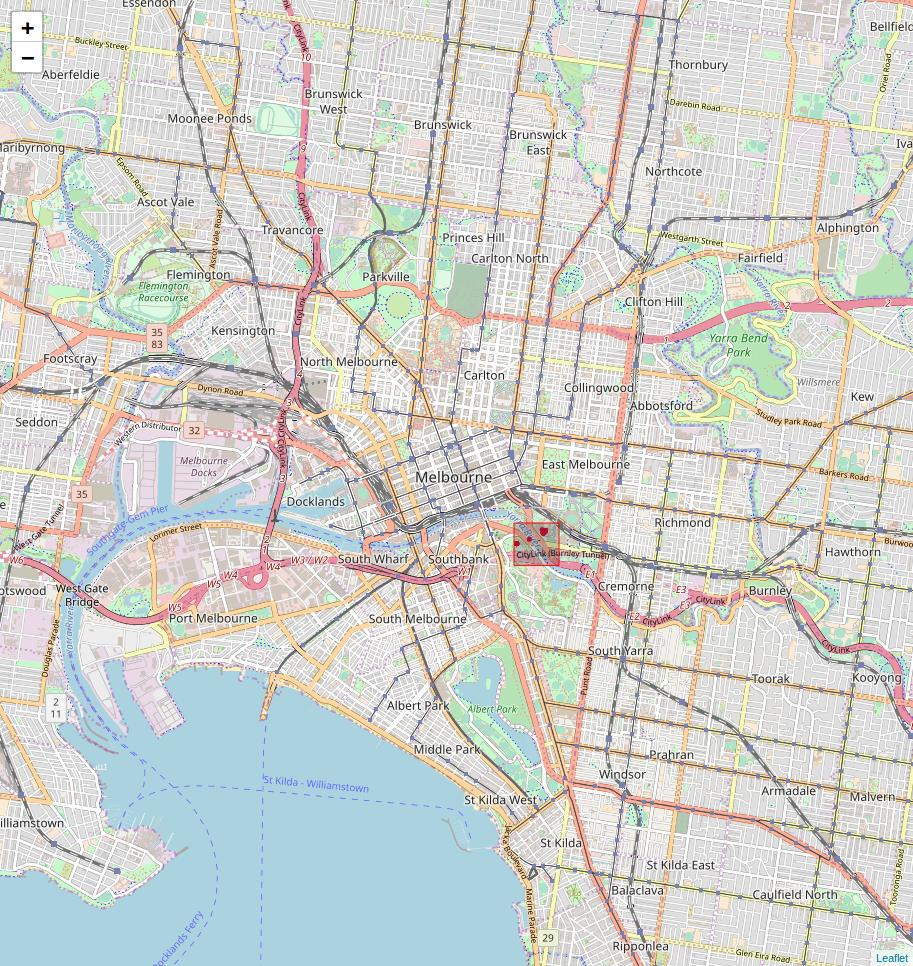}  } 
		&\multicolumn{2}{c}{\includegraphics[trim={15cm 12cm 8cm 15cm},clip,width=0.24\linewidth]{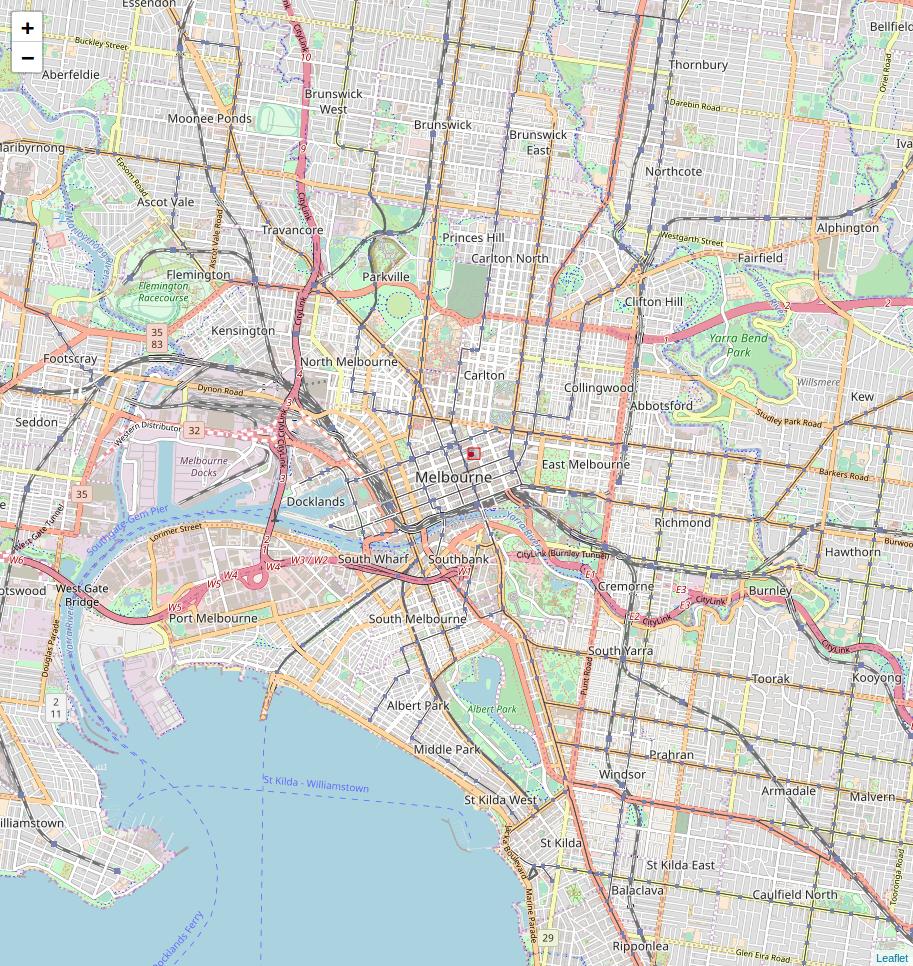}}   
		& 	\multicolumn{2}{c}{\includegraphics[trim={8cm 12cm 15cm 15cm},clip,width=0.24\linewidth]{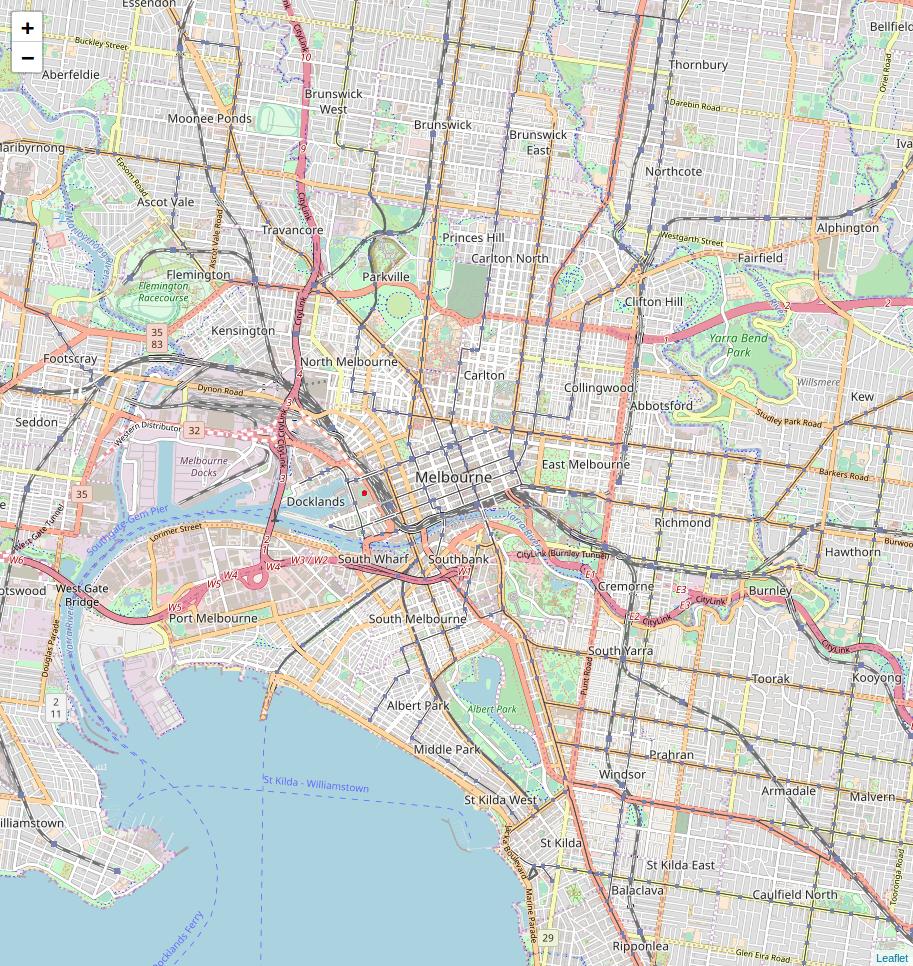}   }
		& \multicolumn{2}{c}{\includegraphics[trim={8cm 12cm 15cm 15cm},clip,width=0.24\linewidth]{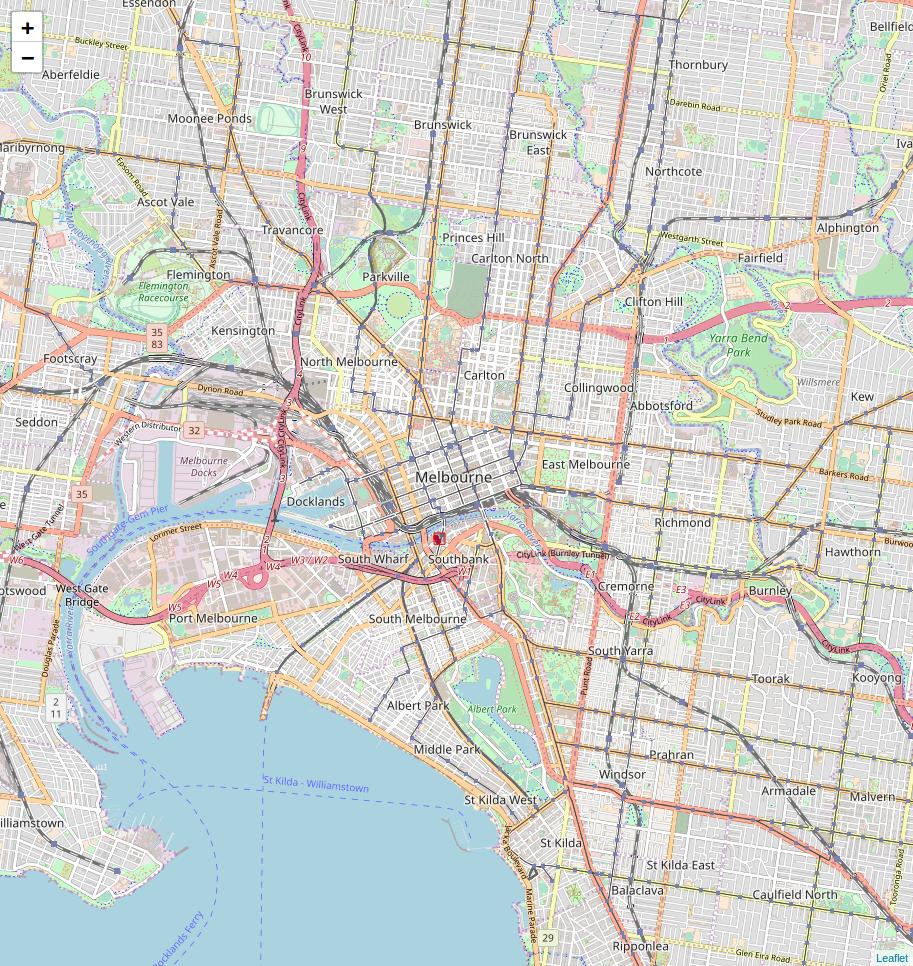}  } \\
		
		Start: 01-28 09:40 &End: 01-28 10:40	&Start: 02-27 00:00 & End: 02-27 01:00 					&Start: 03-10 13:50  &End: 03-10 14:40 						&Start: 04-23 06:00  & End: 04-23 06:50\\
		Tweets count: 35 & Area (sqkm):	0.44	&Tweets count: 17 	& Area (sqkm): 0.028				&Tweets count:  13& Area (sqkm):	0.002 						&Tweets count:  10	& Area (sqkm): 0.028\\
		\textbf{Top 5:} & \#nickcave: 17.1\%& \textbf{Top 5:} & \#greekandthe- city:	82.4\% 				& 	\textbf{Top 5:} & \#purposetour:	46.2\%			& \textbf{Top 5:} & \#tvweeklogies: 50 \%\\
		\#australian- open:	11.4\%	&@serenawilli- ams:	8.6\% 	& \#lovelonsdale:	76.5\% & \#kkfuntimes:	76.5\%  & \#justinbieber:	30.8\%	&\#melbourne:	23.1\% 				& 	\#redcarpet:	20\% & @nazeem:	10\% \\
		\#sidneymyer- musicbowl:	8.6\%& @australian- open:	8.6\%& \#kkrockchic:	64.7\%  & \#mc:	17.6\% 	& \#purposewor- ldtour:	15.4\%& \#belieber:	15.4\%				& @rebeccavall- ance:	10\%  & @hishandso- meself:	10\% \\

		\multicolumn{2}{c}{\includegraphics[trim={12cm 10cm 4cm 11.5cm},clip,width=0.24\linewidth]{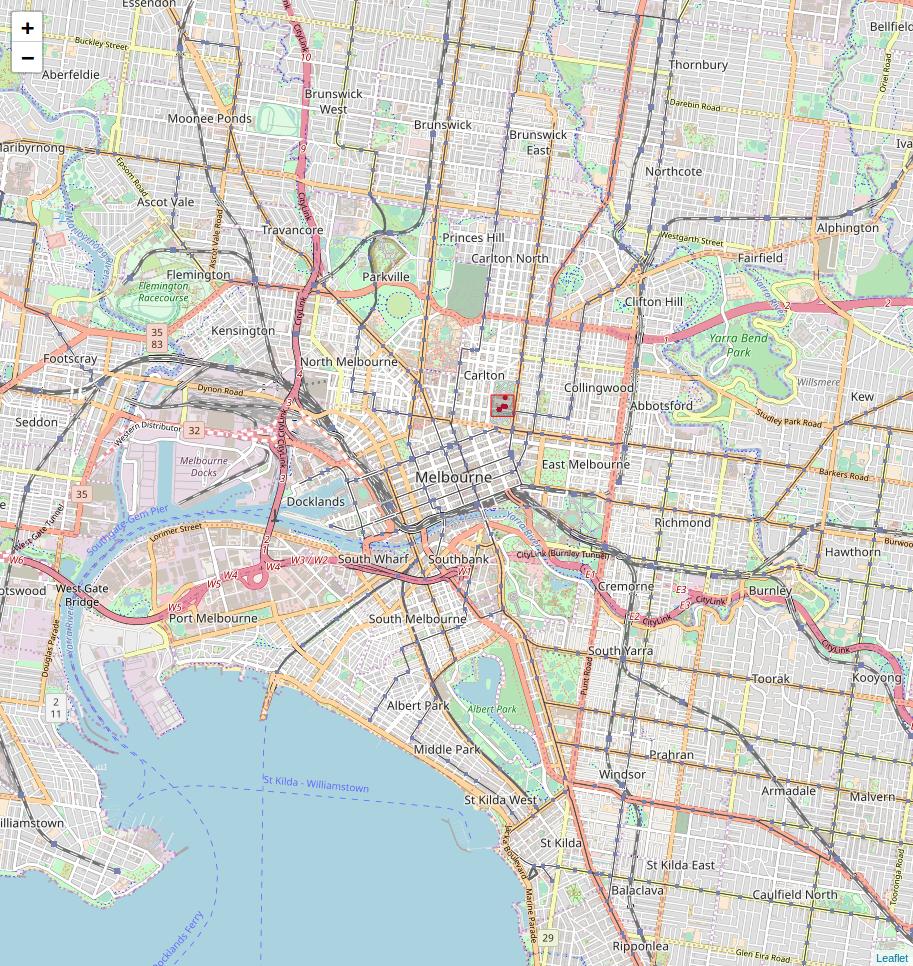} }
		&\multicolumn{2}{c}{\includegraphics[trim={12cm 10cm 4cm 11.5cm},clip,width=0.24\linewidth]{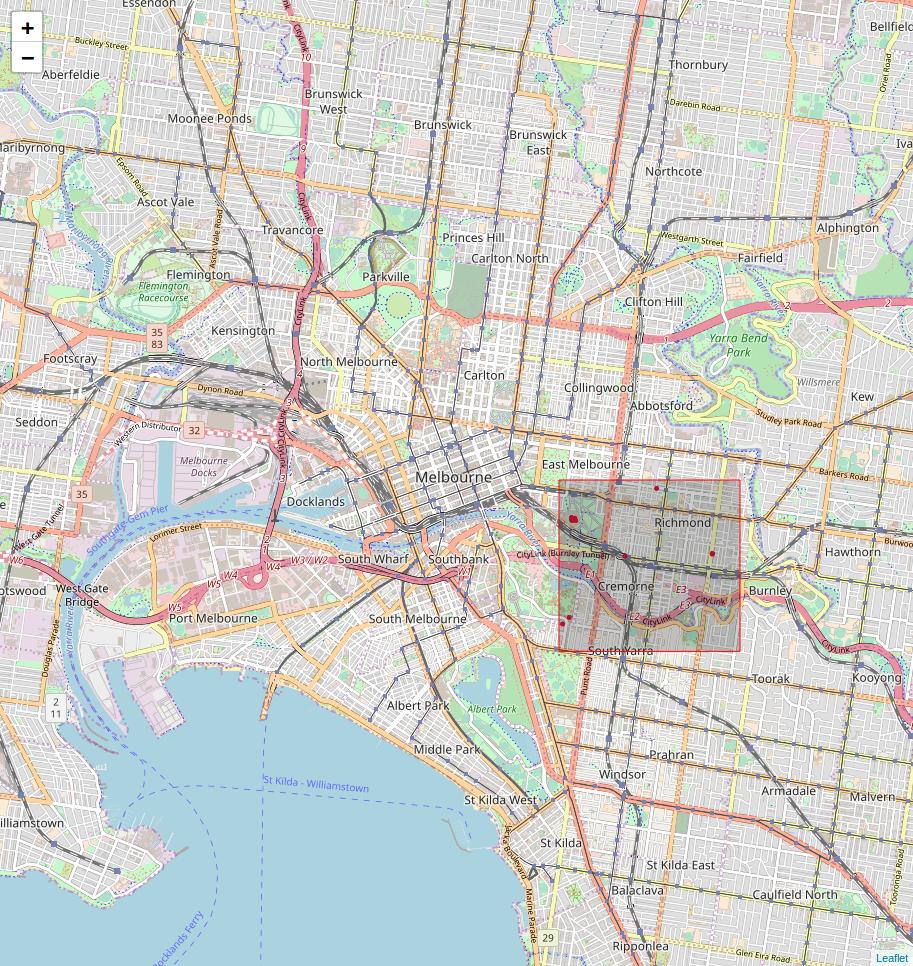} }   
		& \multicolumn{2}{c}{\includegraphics[trim={2cm 14cm 12cm 6cm},clip,width=0.24\linewidth]{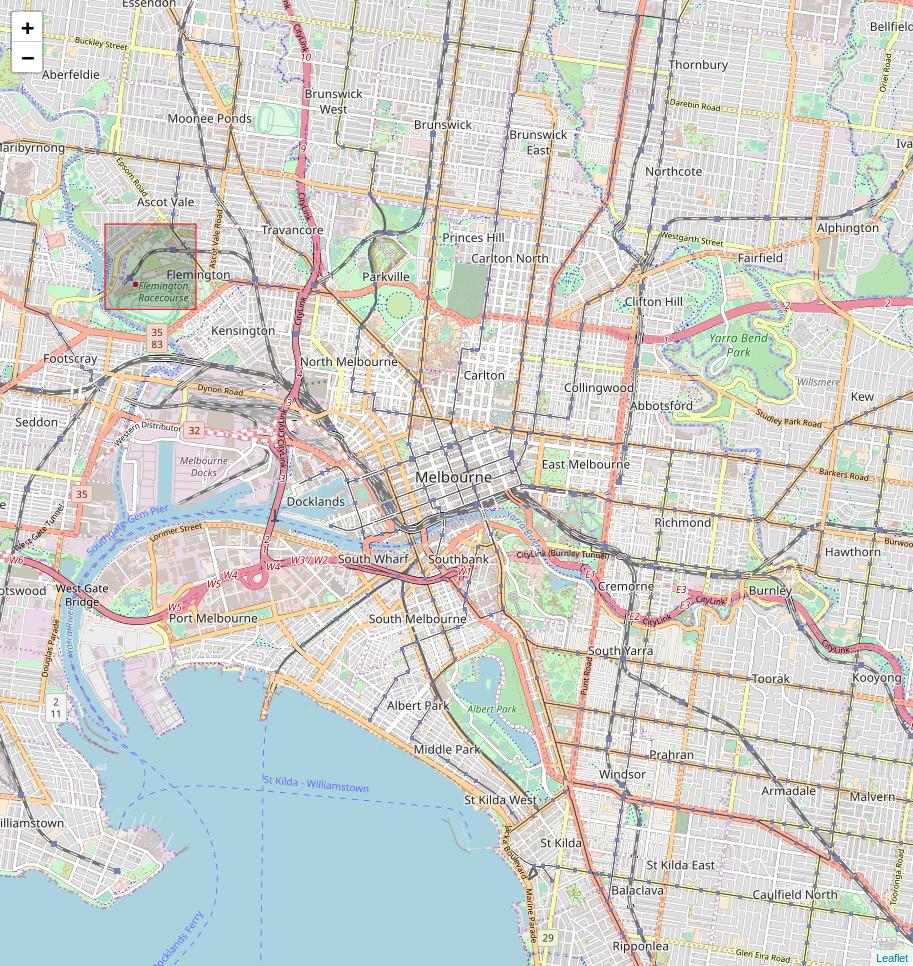}  }		
		&\multicolumn{2}{c}{\includegraphics[trim={8cm 10cm 8cm 11.5cm},clip,width=0.24\linewidth]{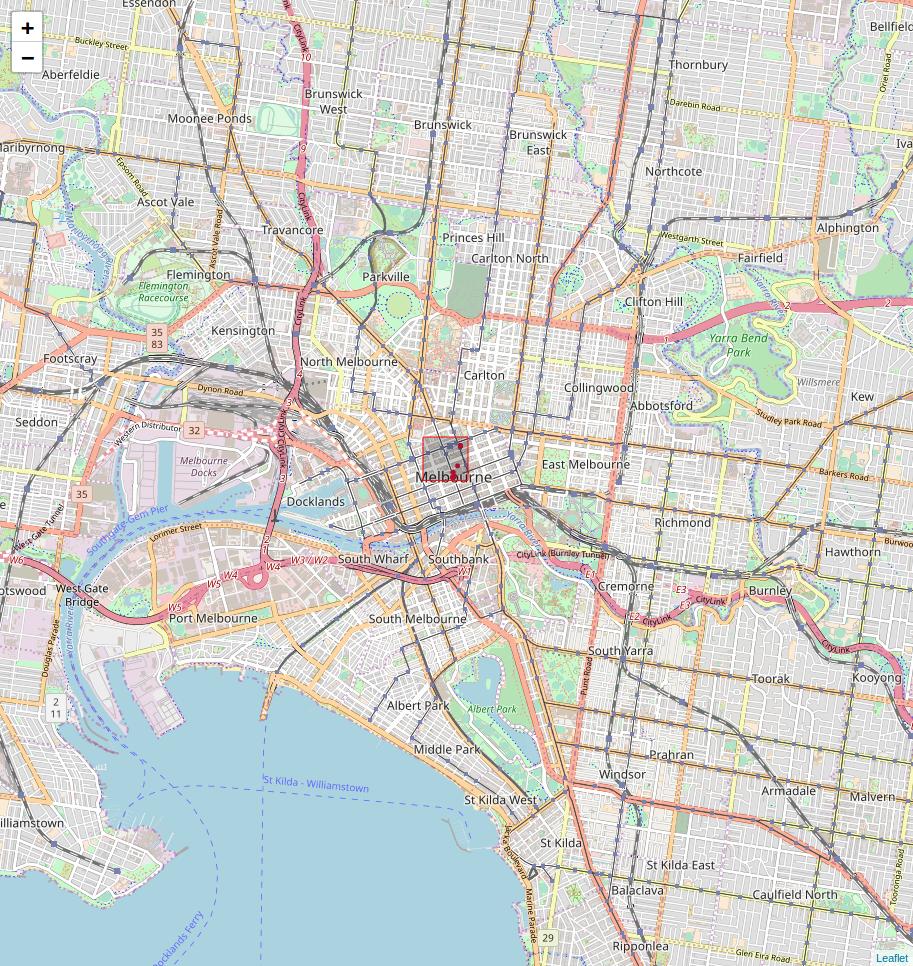}} \\
		
		Start: 05-19 04:00 &End: 05-19 05:10						&Start:  06-25 04:50 & End: 06-25 06:10 				& Start:  10-07 04:20 &End: 	 10-07 05:10			&Start:  11-14 23:00 	 & End: 11-14 23:50\\
		Tweets count: 29 & Area (sqkm):	0.11						&Tweets count: 34  	& Area (sqkm): 7.042 				& Tweets count: 9 & Area (sqkm):	1.761			&Tweets count:  28	& Area (sqkm): 0.44\\
		\textbf{Top 5:} & \#photo:	65.5\%							& \textbf{Top 5:} & \#afl:	29.4\%						&\textbf{Top 5:} & \#winx:	44.4\%						& \textbf{Top 5:} & \#melbourne:	28.6\%\\
		\#gabs2017:	31\%	& @gabsfestival:	13.8\% 				& \#afltigersbl- ues:	14.7\% &\#goblues:	14.7\% 			& \#gowinx:	11.1\%	& \#melbs:	11.1\% 									&\#job:	21.4 \% & \#hiring:	21.4\% \\
		@tcbrewery:	3.4\%& @illiards:	3.4\% 						& \#mcg:	11.8\%  & \#boundbyb- lue:	11.8\% 			& \#spring:	22.2 \%& \#winxxi:	11.1\% 										& \#careerarc:	17.9\%  & \#marriageeq- uality:	14.3\% \\
		
	\end{tabular}
	\caption{Sample detected events using Twitter Data in Melbourne, 2017}
	\label{fig:twitter_results}
	
\end{figure}

To evaluate the performance and reliability of the proposed method, we experiment with the whole dataset for Melbourne in 2017. Figure~\ref{fig:twitter_results} visualises some of the detected events on the map. Each event has the start and end time, total tweets, area and the top 5 hashtags/mentions. We use the top hashtags/mentions along with the event time to manually evaluate the correctness of the event. Table~\ref{tab:flickr_twitter_details}, column "Twitter" shows the total number of flagged events after each phase of the proposed method. In total, we detect 137 events after the removal of all false positives.

\begin{table}[!th]
	\centering
	\caption{Precision results: Melbourne detected events in 2017 using geotagged tweets (random selection of 45 events)}
	\small
	\label{tab:precision_results}
	\setlength\tabcolsep{3pt}
	\setlength\extrarowheight{1pt}
	\resizebox{\linewidth}{!}{
	\begin{tabular}{cm{2.2cm}m{.35\linewidth}ccccm{.35\linewidth}}
		\hline
		\multicolumn{5}{c}{\textbf{Detected Events Results}}&  \multicolumn{3}{c}{\textbf{Manual Assessment}} \\ \hline
		\textbf{ID} &\textbf{ Date \& Time} & \textbf{Top 5 hashtags/mentions: \# of tweets } & \textbf{C} & \textbf{A} & \textbf{SI} & \textbf{?}  & \textbf{Event Description}  \\ \hline
		5 & -S 01-28 09:40  \newline -E 01-28 10:40 &   \#nickcave: 6, \#australianopen: 4, \#23: 4, @serenawilliams: 3, @australianopen: 3 & 35 & 0.44 & 0.57 & 1 & Serena Williams wins Australian Open final (Tennis) \\ \hline
		9 & -S 02-27 00:10  \newline -E 02-27 01:00 &   \#greekandthecity: 13, \#lovelonsdale: 13, \#kkfuntimes: 13, \#kkrockchic: 11, \#mc: 3 & 18 & 0.44 & 2.94 & 1 & Thousands turn out to celebrate Lonsdale St Festival’s 30th anniversary \\ \hline
		
		12 & -S 03-18 06:50  \newline -E 03-18 07:40 &   \#adele: 10, \#melbourne: 4, \#etihadstadium: 3, \#adele2017: 2, \#benandjase: 1 & 14 & 0.44 & 1.43 & 1 & Adele adds new Melbourne show to 2017 Australian tour \\ \hline
		22 & -S 04-05 10:00  \newline -E 04-05 12:20 &   \#worlds50best: 41, @theworlds50best: 28, @australia: 14, \#seeaustralia: 11, \#restaurantaustralia: 10 & 94 & 0.03 & 1.11 & 1 & The 2017 awards ceremony for the World's 50 Best Restaurants 2017 \\ \hline
		26 & -S 04-25 04:00  \newline -E 04-25 05:40 &   \#anzacday: 15, \#lestweforget: 8, @mcg: 8, \#mcg: 6, \#afldonspies: 5 & 44 & 0.01 & 0.95 & 1 & {\color{blue}* Melbourne Cricket Ground: AFL -round 5 - Essendon VS Collingwood. * ANZAC Day public holiday} \\ \hline
		30 & -S 05-21 04:20  \newline -E 05-21 05:50 &   @gabsfestival: 31, \#photo: 26, \#gabs2017: 13, @garage: 8, @stockadebrewco: 2 & 36 & 0.03 & 2.22 & 1 & GABS 2017 global festival: seventy five thousand attendees from across Australia and overseas \\ \hline
		34 & -S 06-04 02:40  \newline -E 06-04 03:30 &   \#dariusisbaptised: 7, \#dariusisone: 6, \#sentimentalsundays: 4, @gozwift: 1, \#ergo: 1 & 12 & 7.04 & 1.58 & 1 & Private event: baptism \\ \hline
		36 & -S 06-09 09:20  \newline -E 06-09 11:00 &   \#bravarg: 10, \#mcg: 8, @mcg: 5, \#football: 5, \#soccer: 3 & 38 & 0.03 & 0.82 & 1 & Soccer game: Brazil VS Argentina \\ \hline
		37 & -S 07-01 07:00  \newline -E 07-01 08:00 &   \#cosplay: 11, \#ozcomiccon: 10, \#doctorwho: 2, \#stormtrooper: 1, @missmelmacklin: 1 & 20 & 0.11 & 1.25 & 1 & Oz Comic-Con event: Showcasing all the latest studio activations, comics, anime, cosplay, and video games \\ \hline
		38 & -S 07-15 04:00  \newline -E 07-15 04:50 &   @gold: 14, @stkildafc: 14, \#hiik: 9, @fbi: 5& 15 & 0.11 & 2.8 & 1 & St Kilda Football Club in support of Marriage Equality \\ \hline
		39 & -S 08-13 23:30  \newline -E 08-14 00:20 &   \#melbournelife: 10, \#melbournecbd: 10, \#melbournecentral: 9, \#hokusai: 7, \#hokusia: 1 & 10 & 0.44 & 3.7 & 1 & National Gallery of Victoria: Hokusai Exhibition \\ \hline
		40 & -S 09-21 03:20  \newline -E 09-21 04:20 &   @shortstopmelb: 2, @bentoonzc: 1, \#sales: 1, \#job: 1, \#melbourne: 1 & 64 & 0.44 & 0.09 & 0 & {\color{red}Not a valid event: General tweets ***} \\ \hline
		\multicolumn{8}{l}{C: Count, A: Area, ?: Event label }
	\end{tabular}
	}
\end{table}

\subsubsection {\textbf{Precision and Strength Index Results}}
We select a random 45 events as the evaluation set. Table~\ref{tab:precision_results} reports the manual evaluation results for sample selected events from the evaluation set. For each event, the algorithm returns the results of the event start and end date-time, area/region in $sqkm$, tweets count, top 5 hashtags/mentions with its occurrences. In total, 40 out of 45 are correct events according to the manual evaluation, with a precision measure of 89\%. False positives are highlighted in red in Table~\ref{tab:precision_results}. The table shows that both local and global events are detected using the proposed method (see the "area" column in Table~\ref{tab:precision_results}), concurrent events such as events \# 26 and 45 (highlighted in blue). The method also finds private events such as event \#34 and opinion events such as event \#38 about marriage equality.

SI index is also reported for all events (see column "SI" in Table~\ref{tab:precision_results}). We obtain SI with an average of 1.2 across the evaluation set. This is an indication that all tweets for an event contain at least one of the relevant hashtags/mentions, which confirms the accurate results of the proposed method.

\subsubsection {\textbf{Recall and Strength Index Results}}
We select 15 events to assess the recall with a total of 10 correctly detected events according to the manual evaluation, with a recall measure of 66.7\% and average SI of 1.034. The reduction in recall is explainable since the social media does not contain information about all actual events. This causes a certain increase in the false negatives. Table~\ref{tab:recall_results} reports the date-time, top 5 entities, tweets count, event area and SI for sample selected events from the evaluation set. In the table, the false negatives are highlighted in blue. 

\begin{table}[!th]
	\centering
	\caption{Recall results: Melbourne common events occurred in 2017}
	\small
	\label{tab:recall_results}
	\setlength\tabcolsep{3pt}
	\setlength\extrarowheight{3pt}

	\begin{tabular}{cm{3cm}m{2.1cm}m{5cm}ccc}
		\hline\hline
		
		\textbf{\#} & \textbf{Event Description} & \textbf{ Date-time} & \multicolumn{1}{c}{ \textbf{Entity: occurrences }} & \textbf{*} & \textbf{Area} & \textbf{SI}  \\ \hline\hline
		1&The Night Market; Wed 6-9pm
		&-S 01-18 10:10  -E 01-18 11:00& \#melbourne: 3,  \#queenvictoriamarket: 2, \#glutenfree: 1, \#nofilter: 2, @hunde: 1 & 35&28.181&0.258\\ \hline
		
		2&Moomba Festival;  10-13 Mar 
		&-S 03-12 01:30  -E 03-12 01:50& \#melbourne: 4,  \#davidhockney: 1, \#shrineofremembrancemelbourne: 1, \#ladiesinblack: 1, @moombafestival: 1&16&7.04&0.498\\ \hline
		5&Patti Smith performs; Apr 16&-S 04-16 13:20  -E 04-16 14:00& \#music: 4,  \#livemusic: 4, \#melbourne: 4, \#horses: 3, \#pattismith: 3 & 8&0.028&2.25\\ \hline
		
		6&Australia Day; Jan 26&-S 01-26 05:40  -E 01-26 06:10& \#art: 6,  @ngv: 5, \#ipad: 5, @australianopen: 5, \#inspiration: 5& 20&1.76&1.3\\ \hline
		
		7&Labour Day; Mar 13& \multicolumn{5}{c}{ {\color{blue} \textit{Not Detected} }}\\ \hline
		9&Easter Sunday; Apr 16&-S 04-17 04:00  -E 04-17 04:20& \#brunch: 1,  \#eastersunday: 1 &4&7.04&0.5\\ \hline
		
		10&ANZAC Day; Apr 25&-S 04-25 04:00  -E 04-25 05:40& \#anzacday: 15,  \#lestweforget: 8, @mcg: 8, \#mcg: 6, \#afldonspies: 5&44&0.007&0.96\\ \hline
		
		11&Queen's Birthday; Jun 12& \multicolumn{5}{c}{ {\color{blue} \textit{Not Detected} }}\\ \hline
		13&Melbourne Cup; Nov 7&-S 11-07 06:40  \newline -E 11-07 07:10& \#melbournecup: 5,  \#foodlover: 1, \#burger: 1, \#foodie: 1, \#melbourne: 5 &11&1.76&1.183\\ \hline
		15&Boxing Day; Dec 26&-S 12-26 02:00  \newline -E 12-26 02:40& \#boxingdaytest: 6,  \#beatengland: 5, \#mcg: 4, \#ashes2017: 3, \#kneipping: 3&10&0.007&2.1\\ \hline \hline
		
	\end{tabular}
\end{table}

\subsection{Case Study: Flickr Dataset}
\label{sec:res-caseStudyFlickr}
In this section, we present our second case study for event detection using the Flickr dataset introduced in Section~\ref{sec:dataset}. The dataset was further reduced by keeping only geotagged images for January-2013 for Melbourne and New York, while for London and Paris we selected data for January 2012. The chosen years have largest number of photos taken in January. We evaluated the proposed event detection method using the set of images collected for each city. In our experiments, instead of extracting top k hashtags/mentions as in the twitter case study, we use the title, user tags and description attributes for each image to extract the most frequent entities of an event. Table \ref{tab:flickr_twitter_details}, column "Flickr" reports the results each individual phase of the proposed method using the Flick dataset for each city. 

\begin{figure}[!th]
	\centering
	\scriptsize
	\setlength\tabcolsep{1pt}
	\setlength\extrarowheight{0pt}
	\begin{tabular}{p{0.245\linewidth}p{0.245\linewidth}p{0.245\linewidth}p{0.245\linewidth}}
		
		\multicolumn{2}{c}{
			\includegraphics[trim={7cm 9cm 12cm 15cm},clip,width=.49\linewidth]{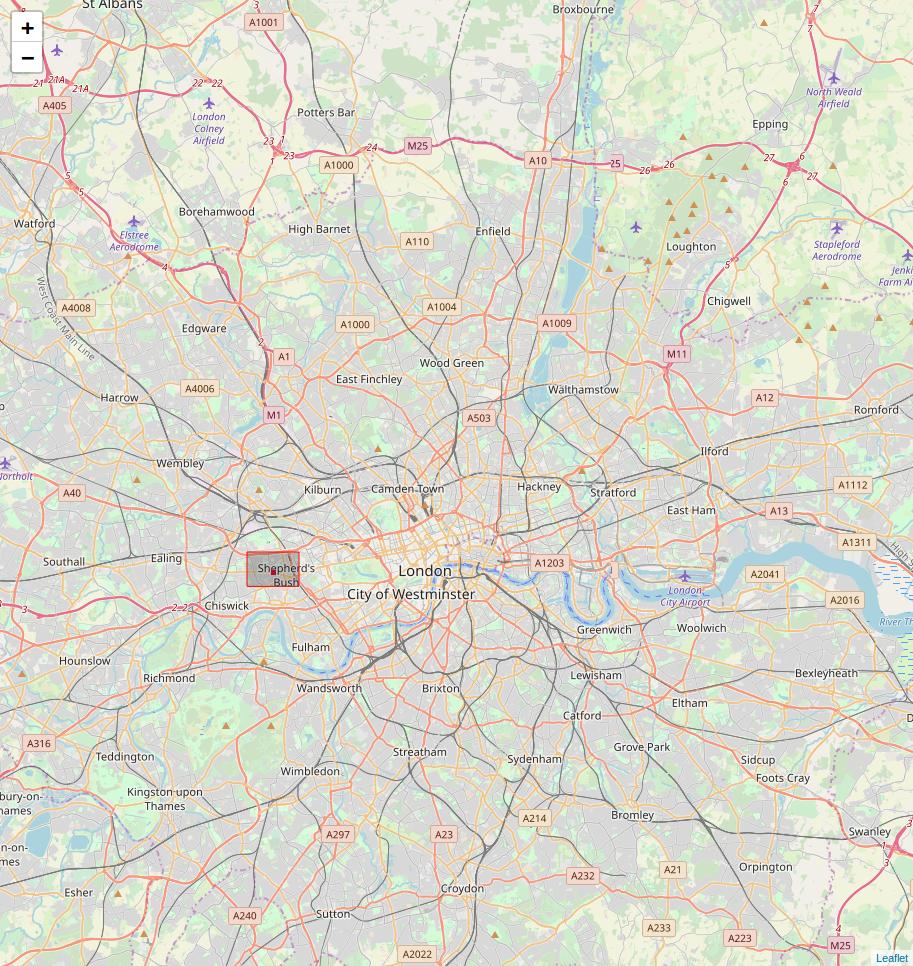}} & 
		\multicolumn{2}{c}{\includegraphics[trim={7cm 9cm 12cm 15cm},clip,width=.49\linewidth]{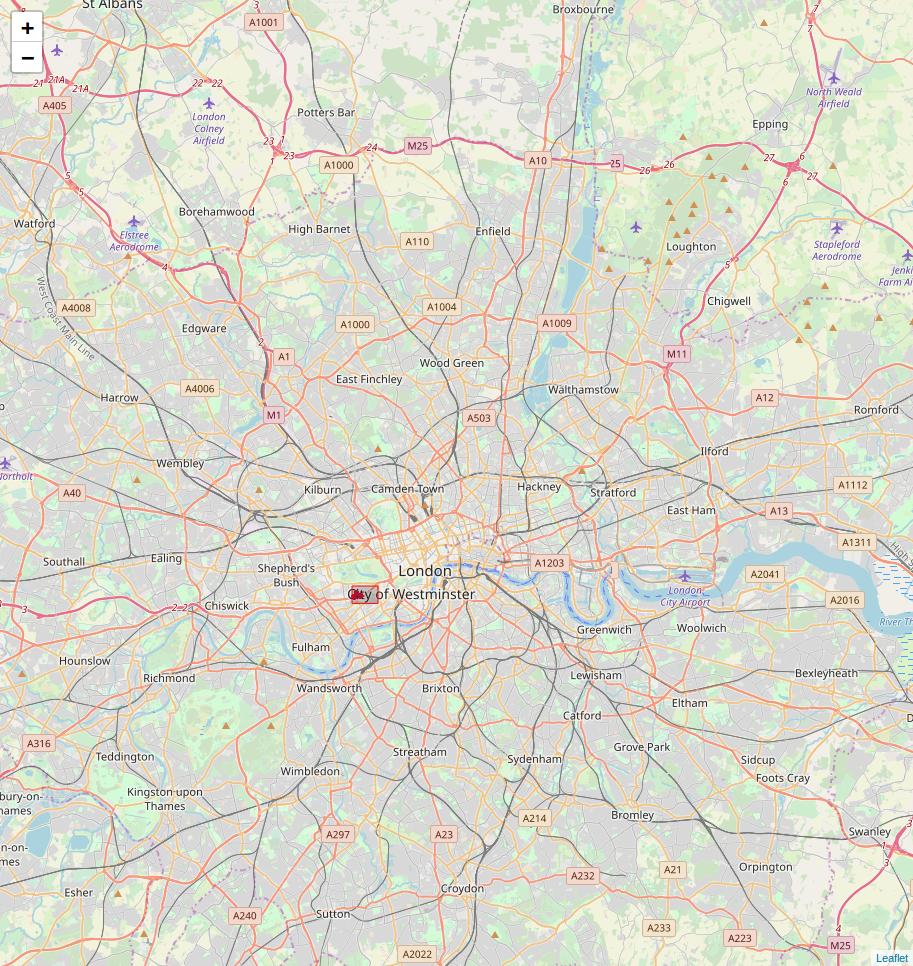} } 
		\\
		
		Start: 01-20 22:50 &End: 01-21 00:20	&Start: 01-25 14:50 & End: 01-25 16:50\\
		Images count: 40 & Area (sqkm):	4.089	&Images count: 82 	& Area (sqkm): 1.023\\
		\textbf{Top 5:} & danny:	7\%			& \textbf{Top 5:} & museum:	19.6\% \\
		rock:	6.9\%	&ben:	6.9\% 			& science:	10.3\% & gallery:	9.9\% \\
		bowes:	3.9	\%& bush:	3.6	\%			& art:	9.8\%  & london:	9.8\% \\
		SI = 0.283 && SI = 0.594 & \\

		\multicolumn{2}{p{.49\linewidth}}{Event: An evening with Danny \& Ben from Thunder, London-England, 20 January,2012} &
		\multicolumn{2}{p{.49\linewidth}}{Event: Howler performing at Rough Trade East, Brick Lane, London on 25-January 2012} \\
		
		\multicolumn{4}{c}{		 \text{(a) } } \\
		
		\multicolumn{2}{c}{\includegraphics[trim={12cm 10.2cm 8cm 15cm},clip,width=.49\linewidth]{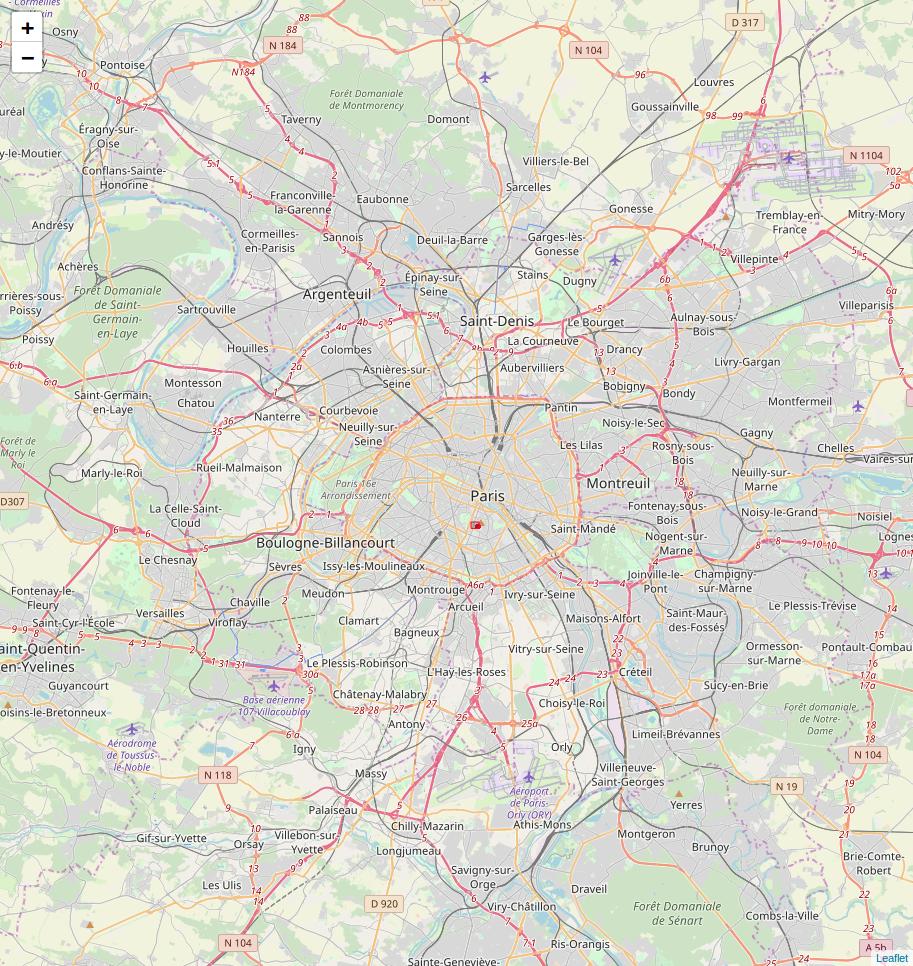}}
		& \multicolumn{2}{c}{\includegraphics[trim={12cm 10.2cm 8cm 15cm},clip,width=.49\linewidth]{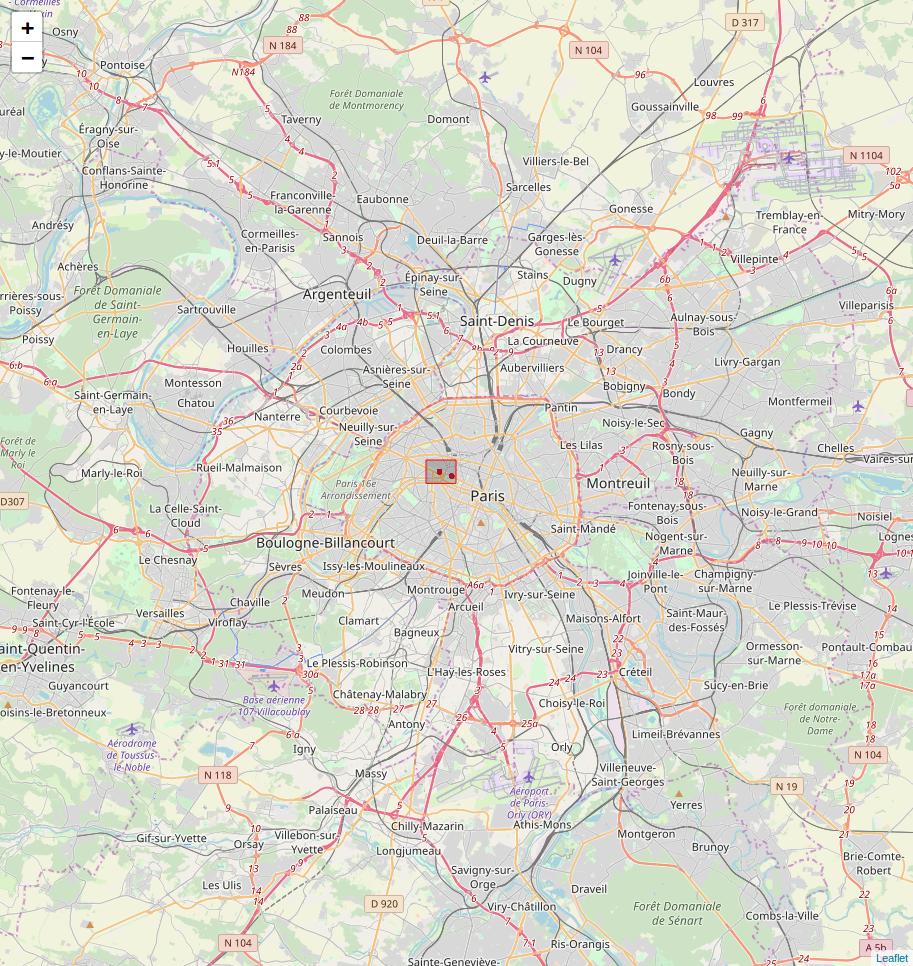}}  \\
		
		Start: 01-09 11:00  &End: 01-09 11:50	&Start: 01-28 13:40  & End: 01-28 16:20\\
		Images count: 16 & Area (sqkm):	0.108	&Images count: 82 	& Area (sqkm): 1.723\\
		\textbf{Top 5:} & workshop:	12.5\%			& \textbf{Top 5:} & tedxconcorde:	19.4\% \\
		arduino:	12.5\%	&ensad:	12.5\% 			& paris:	15.1\% & tedx:	9.7\% \\
		day:	6.2	\%& lab:	6.2	\%			&3a:	8\%  & france:	5.5\% \\
		
		SI = 0.499 && SI = 0.577 &\\
		
		\multicolumn{2}{p{.49\linewidth}}{Event: Arduino workshop, day one, Paris-France, 9 January, 2012} & 
		\multicolumn{2}{p{.49\linewidth}}{Event: TEDx Concorde event in Paris, January 28, 2012 - Theme La Diversite} \\
		\multicolumn{4}{c}{		 \text{(b) } } 
	\end{tabular}
	
	\caption{Sample event detection results using Flickr images, January-2012 in: (a) London, (b) Paris}
	\label{fig:flickr_results_2012}
	
\end{figure}

Figure \ref{fig:flickr_results_2012} shows a sample of detected events in London and Paris, January-2012 using the Flickr dataset. We manually evaluate the detected events by using Google and looking at the tags, description and title of images in an event. The figure includes the description of each event as well as the computed SI index. 
Similarly, we were able to identify and verify various detected events in Melbourne and New York for January-2013.
These results demonstrate the effectiveness of the proposed method for event detection in different social media.

\subsection{Comparative Analysis of Baseline Algorithms}
\subsubsection{POI-based Event Detection}

\label{sec:res-compResults}
We use the January-2017 Melbourne geotagged tweets dataset to run this experiment. We selected 242 POIs (100m x 100m) in Melbourne where most of events occur. We apply the proposed event detection algorithm on the POIs instead of the the multiscale grid generated by quad-tree method. Every $\Delta T$ (i.e. 10-min), we flag all POIs with smoothed signal less than the threshold ($\tau_2$). 

\begin{table}[!th]
	\centering
	\caption{Melbourne detected events in 28-January 2017 using each of POI and Quad-tree methods}
	\small
	\label{tab:poi_results}
	\setlength\tabcolsep{3pt}
	\setlength\extrarowheight{3pt}
	\begin{tabular}{ccm{0.6\linewidth}cc}
		\hline\hline
		{\textbf{\#}} & \textbf{Time-Duration} &  {\textbf{Top 5 hashtags/mentions} (hashtag/mention : \% of occurrences)} & \textbf{Count}  &\textbf{Area} \\ \hline\hline
		\multicolumn{5}{c}{\textbf{POI events}} \\ \hline
		
		1 &  -S 08:30 -D 40 & \#ausopen: 37.5, @australianopen: 25, \#williamssisters: 18.8, @serenawilliams: 12.5, \#melbourne: 12.5 & 16 & 0.01 \\ 
		2 & -S 10:00  -D 20 & \#nickcave: 80, \#sidneymyermusicbowl: 60, \#livemusic: 20, \#melbourne: 20 & 5 & 0.01 \\ 
		3 & -S 11:10  -D 40 & \#ausopen: 50, \#womensfinal: 33.3, \#rodlaverarena: 33.3, \#ausralianopen: 25, \#serena: 25 & 12 & 0.01 \\ \hline 
		\multicolumn{5}{c}{\textbf{Quad-tree events}} \\ \hline	
		1 & -S 08:30  -D 40 & \#ausopen: 33.3, \#williamssisters: 23.8, @australianopen: 23.8, @venuseswilliams: 14.3, @serenawilliams: 14.3 & 21 & 0.052 \\ 
		2 & -S 08:40  -D 30 & \#ausopen: 50, \#williamssisters: 25, @australianopen: 25, \#melbourne: 16.7, \#grandslam: 8.3 & 12 & 0.003 \\ 
		3 & -S 10:00  -D 20 & \#nickcave: 80, \#sidneymyermusicbowl: 60, \#livemusic: 20, \#melbourne: 20 & 5 & 0.013 \\ 
		4 & -S 10:00  -D 40 & \#nickcave: 20.8, @serenawilliams: 12.5, \#23: 16.7, \#australianopen: 12.5, \#sidneymyermusicbowl: 12.5 & 24 & 0.834 \\ 
		5 & -S 10:00   -D 50 & \#nickcave: 13.2, @serenawilliams: 10.5, \#sidneymyermusicbowl: 7.9, \#23: 10.5, \#australianopen: 7.9 & 38 & 13.347 \\ 
		6 & -S 10:20  -D 20 & @australianopen: 28.6, \#serenavsvenus: 14.3, \#23: 28.6, \#venus: 14.3, \#williamssisters: 14.3 & 7 & 0.003 \\ 
		7 & -S 11:10  -D 30 & \#ausopen: 54.5, \#womensfinal: 36.4, \#rodlaverarena: 36.4, \#ausralianopen: 27.3, \#serena: 27.3 & 11 & 0.003 \\ 
		8 & -S 20:50  -D 20 & \#melbourne: 100, \#hiring: 53.8, \#job: 53.8, \#bourkestreet: 46.2, \#careerarc: 46.2 & 13 & 0.052 \\ \hline\hline 
		
	\end{tabular}
\end{table}

Table~\ref{tab:poi_results} reports the details of the detected events on $28/01/2017$ using each of the proposed and baseline methods. 
The results show that our approach is able to detect all events that were identified by the baseline method, along with additional events with different spatial scales that the baseline method was unable to detect. Also, the conducted experiments show that POI-based method detects events with shorter duration than the quad-tree based method. The reason is that POI grid cells are small (100m x 100m).

\begin{figure} [!th]
	\centering
	\setlength\tabcolsep{1pt}
	\setlength\extrarowheight{0pt}
	\resizebox{\linewidth}{!}{
	\begin{tabular}{p{3.7cm}p{3.7cm}p{3.7cm}p{3.7cm}}
		\multicolumn{2}{c}{\includegraphics[trim={15cm 10cm 15cm  10cm},clip,width=7cm]{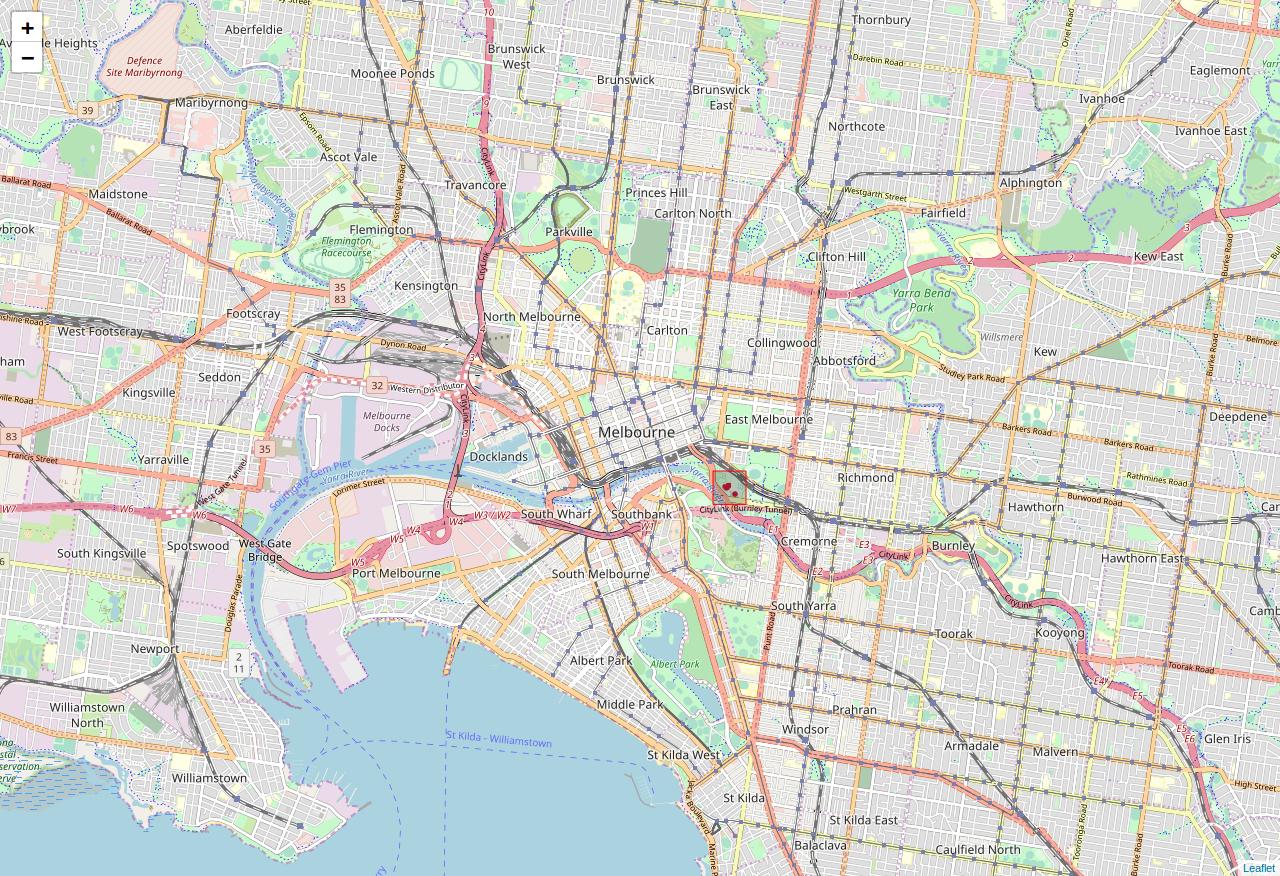}  } &
		\multicolumn{2}{c}{\includegraphics[trim={15cm 10cm 15cm  10cm},clip,width=7cm]{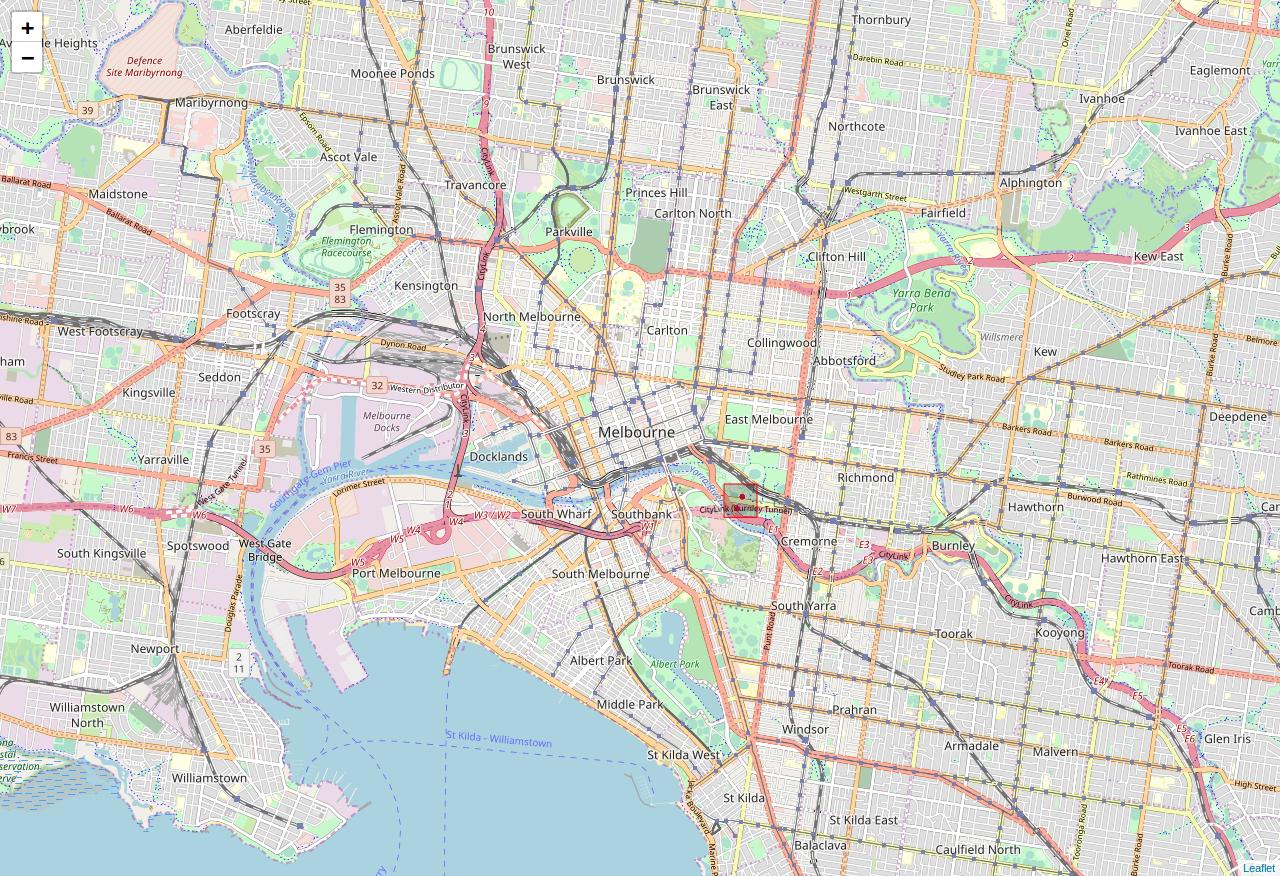}  } \\
		
		Start: 01-14 08:20 &End: 01-14 11:10			&Start: 02-24 5:50 & End: 02-24 8:40				\\
		Tweets count: 11 & SI:	1.636					&Tweets count: 7 	& SI:	 1.571						\\
		\textbf{Top 5:} & \#tennis: 5					& \textbf{Top 5:} & \#melbwc17:	4 					\\
		\#ausopen:	5 &\#australianopen:	4 			& \#gymnasticsworldcup:	3 & \#polskiprogram:	2  	 \\
		\#sport:2& \#melbourne:	2						& @streetsmithmelb:	1  & \#melbourne:	1 	 \\
		
		\multicolumn{2}{c}{\includegraphics[trim={15cm 10cm 15cm  10cm},clip,width=7cm]{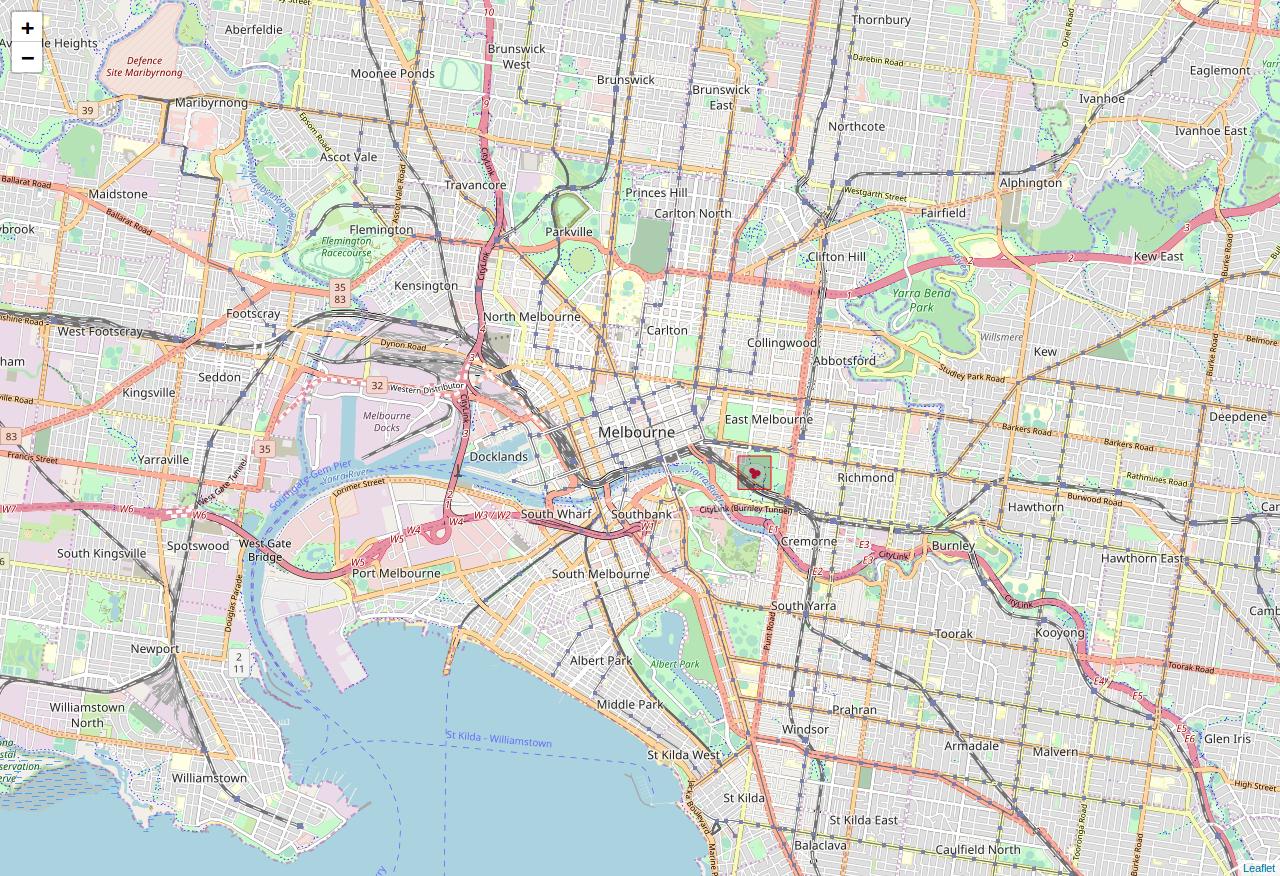}  } &
		\multicolumn{2}{c}{\includegraphics[trim={15cm 10cm 15cm  10cm},clip,width=7cm]{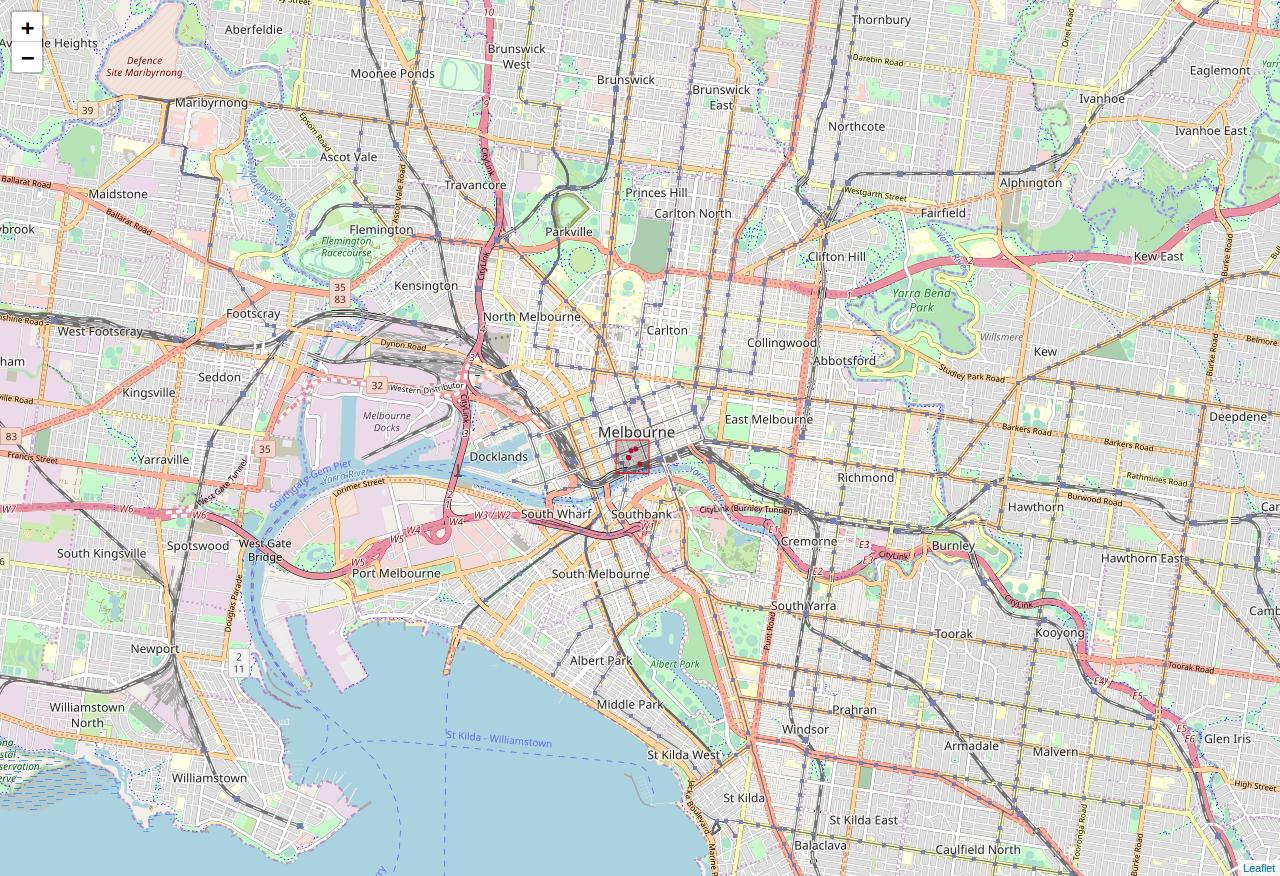}  } \\
		
		Start: 04-23 5:20  &End: 04-23 8:00 				&Start: 05-05 06:00  & End: 21-20 0:10\\
		Tweets count:  14& SI:	0.786 						&Tweets count:  6	& SI: 0.833\\
		\textbf{Top 5:} & @mcg:	5						& \textbf{Top 5:} & @golfclearanceau: 1\\
		@hawthornfc:	3	&\#alwayshawthorn: 1			& 	\#allstarcomics:	1 &\#melbourne:	1 \\
		\#wow:	1& \#strongasone:	1						& \#temellijewellery:	&\#fcbd	:	1 \\

		\multicolumn{2}{c}{\includegraphics[trim={15cm 10cm 15cm  10cm},clip,width=7cm]{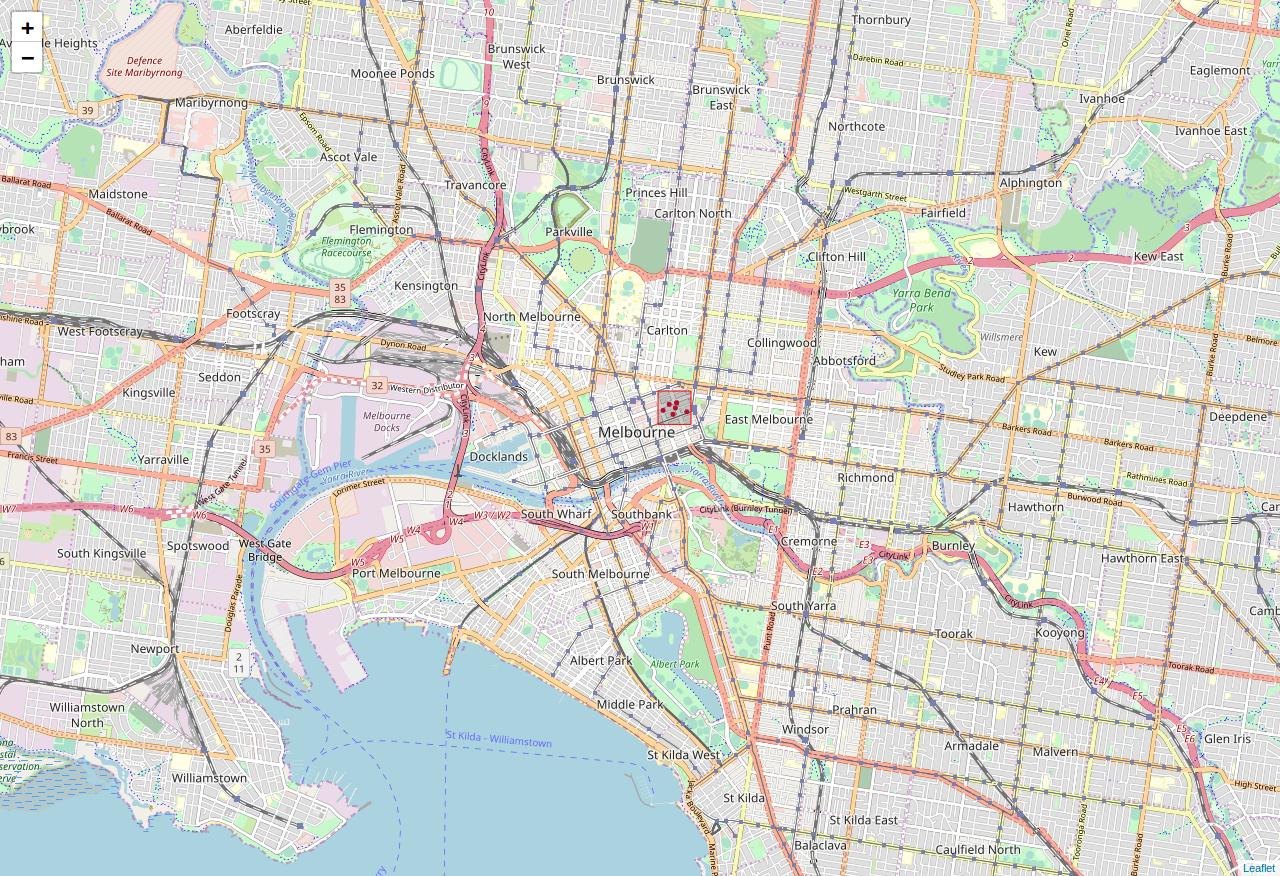}  } &
		\multicolumn{2}{c}{\includegraphics[trim={15cm 10cm 15cm  10cm},clip,width=7cm]{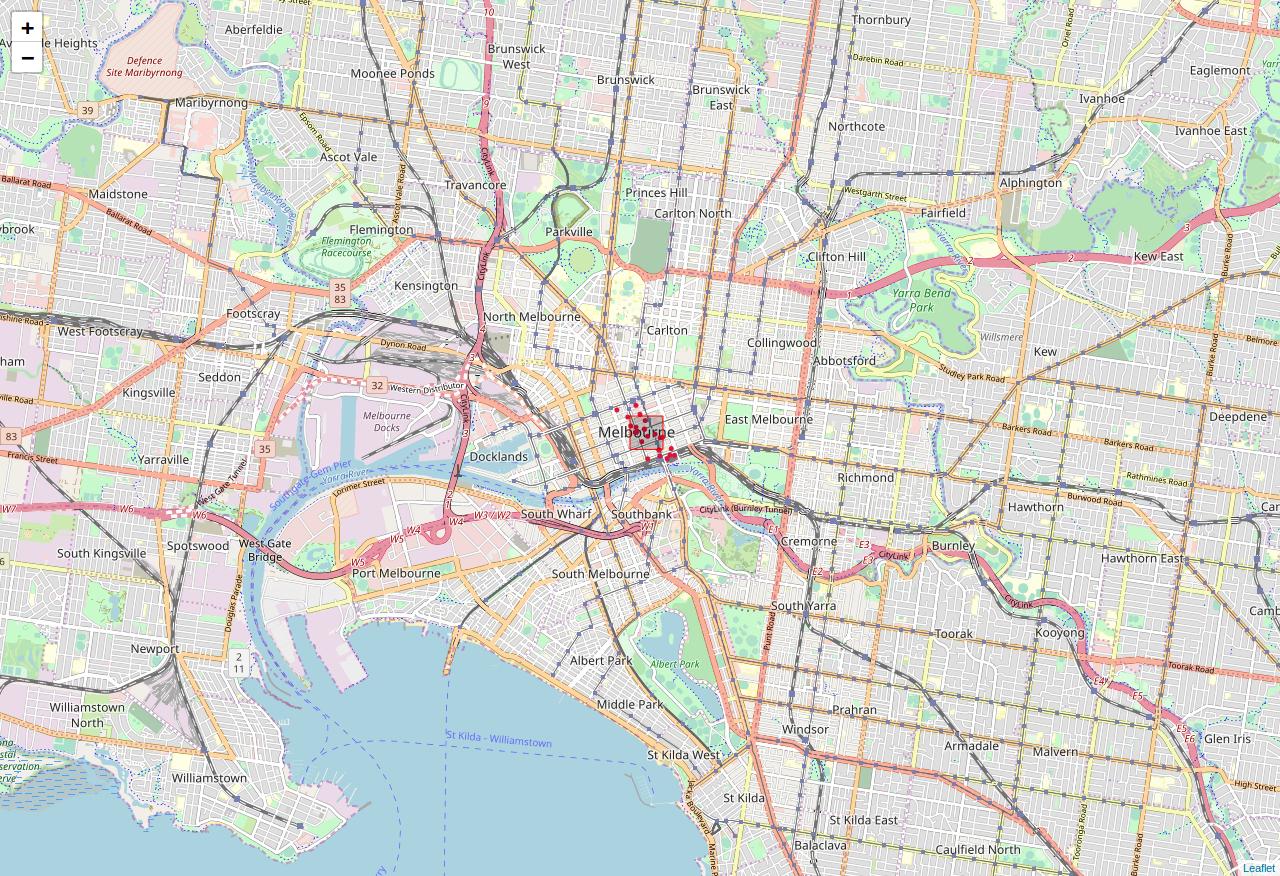}  } \\ 
		
		Start: 07-23 04:20 &End: 07-23 7:10			&Start: 12-30 23:20 & End: 12-31 2:10				\\
		Tweets count: 7 & SI:	0.857					&Tweets count: 28 	& SI:	 0.536						\\
		\textbf{Top 5:} & \#peterpan: 2					& \textbf{Top 5:} & \#melbourne:	8 					\\
		\#wheninmelbourne:	1 &\#street:	1 			& \#job:	3 & \#culturaltart:	2  	 \\
		\#pirates:1& \#twerk:	1						& \#money:	1  & @acmi:	1 	 
	\end{tabular}
	}
	\vspace{-3ex}	\caption{Clustering based event detection results using Twitter Data in Melbourne, 2017}
	\label{fig:clust_vis_results}
\end{figure}

\subsubsection{Incremental Clustering for Event Detection}
We implemented the full algorithm in Python and experimented using the geotagged tweets dataset for Melbourne in 2017. We run the algorithm using the following parameters: time tick $t=10$min, $R=0.25km$, $\Delta T=3$hrs, $K=2$, and $N=5$. The values of the first two parameters are chosen to be close to the values used in our approach for fair comparison. While the values of the other parameters are chosen to be the same as the ones used in the paper \cite{Andrienko-2015}. 

For each time tick, i.e. 10min, we record the set of significant unions (i.e. clusters). In total, we obtained $154546$ significant events in Melbourne 2017. To avoid detecting same event many times through different time ticks, we group the set of detected events by their union ID. Then we keep only the event with the highest number of tweets for each union ID. This grouping strategy has decreased the total number of significant events to 3994, since it removes all redundant events. Figure \ref{fig:clust_vis_results} visualises some of the detected events on the map. Each event has the start and end time, total tweets, SI index and the top 5 hashtags/mentions. 

For quantitative comparison with our approach, we compute precision and recall for the implemented algorithm. Similar to the evaluation of our approach, we use the top hashtags/mentions along with the event time to manually evaluate the correctness of the event.

\begin{table} [!t]
	\centering
	\caption{Precision results for clustering based event detection: Melbourne detected events in 2017 using geotagged tweets (random selection of 20 events)}
	\scriptsize
	\label{tab:clus_precision_results}
	\setlength\tabcolsep{3pt}
	\setlength\extrarowheight{2pt}
	\begin{tabular}{cm{.15\linewidth}m{.31\linewidth}ccccm{.34\linewidth}}
		\hline\hline
		\multicolumn{5}{c}{\textbf{Detected Events Results}}&  \multicolumn{3}{c}{\textbf{Manual Assessment}} \\ \hline
		\textbf{ID} &\textbf{Date \& Time} & \textbf{Top 5 hashtags/mentions: \# of tweets } & \textbf{C} & \textbf{A} & \textbf{SI} & \textbf{?}  & \textbf{Event Description}  \\ \hline\hline
		1 & -S 01-10 13:00	\newline -E 01-10 13:40 &  & 6 & 0.25 & 0 & 0 & Not a valid event \\ \hline 
		2 & -S 01-14 8:20	\newline -E 01-14 11:10 & \#tennis: 5, \#ausopen: 5, \#australianopen: 4, \#sport: 2, \#melbourne: 2 & 11 & 0.25 & 1.636 & 1 & Australian Open from 10 Jan 17 to 29
		Jan 17. \\ \hline		
		3 & -S 02-24 5:50 \newline -E 02-24 8:40 & \#melbwc17: 4, \#gymnasticsworldcup: 3, \#polskiprogram: 2, @streetsmithmelb: 1, \#melbourne: 1 & 7 & 0.25 & 1.571 & 1 & 2017 World Cup Series at Melbourne's Hisense Arena from  22 Feb 17 to 25 Feb 17. \\ \hline 
		4 & -S 02-26 4:40 \newline -E 02-26 7:30  & \#melbourne: 4, \#viktorandrolf: 2, \#black: 2, \#sketchbook: 2, \#help: 2 & 33 & 0.25 & 0.364 & 1 & National Gallery of Victoria: Dutch fashion designers Viktor \& Rolf are on display from 21 Oct 16 to 26 Feb 17. \\ \hline 
		5 & -S 03-1 21:10 \newline -E 03-2 0:00 & & 5 & 0.25 & 0 & 0 & Not a valid event \\ \hline 		
		6 & -S 03-19 22:20 \newline -E 03-19 23:50& \#firealarm: 1, @estellereport: 1, @vamff: 1& 5 & 0.25 & 0.6 & 0 & Not a valid event  \\ \hline
		
		7 & -S 03-21 1:50 \newline -E 03-21 4:10& @tonydohertyoz: 2, \#selfies: 1, \#figure: 1, \#melbourne: 1, \#southwharf: 1 & 5 & 0.25 & 1.2 & 0 &  Not a valid event \\ \hline 
		8 & -S 	04-2 9:10 \newline -E 04-2 11:50& \#dixiechicks: 2, \#happinessis: 1, \#concertphotography: 1, \#melbourne: 1, \#concert: 1 & 6 & 0.25 & 1 & 1 & Dixie Chicks at Rod Laver Arena on  1 Apr 17 and 2 Apr 17 \\ \hline 
		9 & -S 04-8 23:50 \newline -E 04-9 1:30 & \#r4k: 4, \#dolphinproducts: 1, \#teamhmb: 1, \#mazdar4k: 1, \#melbourne: 1 & 5 & 0.25 & 1.6 & 1 & 2017 Herald Sun/City Link Run for the Kids on 9 Apr 17. \\ \hline 
		10 & -S 04-23 5:20 \newline -E 04-23 8:00& @mcg: 5, @hawthornfc: 3, \#alwayshawthorn: 1, \#wow: 1, \#strongasone: 1 & 14 & 0.25 & 0.786 & 1 & Melbourne Cricket Ground: Hawthorn - West Coast) \\ \hline 
		11 & -S 05-5 21:20 \newline -E 05-6 0:10 & @golfclearanceau: 1, \#allstarcomics: 1, \#melbourne: 1, \#temellijewellery: 1, \#fcbd: 1 & 6 & 0.25 & 0.833 & 1 & Free Comic Book Dayon 6 May 17. \\ \hline 
		12 & -S 06-21 2:10 \newline -E 06-21 2:50& @fowl2: 1, @codeblackcoffee: 1, @blognaturgesetz: 1, @hnrysml: 1, @bronte: 1 & 10 & 0.25 & 0.5 & 0 & Not a valid event \\ \hline 
		13 & -S 07-2 1:30 \newline -E 07-2 4:10& \#vflmagpies: 4, \#gopies: 1 & 7 & 0.25 & 0.714 & 1 & VFL round 11 : Collingwood v Footscray on  2 Jul 17 in Victoria Park. \\ \hline 
		14 & -S 07-19 9:30 \newline -E 07-19 12:10& \#melbourne: 5, \#amritmaan: 1, \#dansperspective: 1, \#film: 1, \#dovetonboymadegood: 1 & 21 & 0.25 & 0.429 & 0 & Not a valid event \\ \hline 
		15 & -S 07-23 4:20 \newline -E 07-23 7:10& \#peterpan: 2, \#wheninmelbourne: 1, \#street: 1, \#pirates: 1, \#twerk: 1 & 7 & 0.25 & 0.857 & 1 & The Adventures of Peter Pan \& Tinker Bell: Comedy Theatre from 23 Jun  to 30 Jul 17. \\ \hline 
		16 & -S 08-20 1:50 \newline -E 08-20 4:10 & \#fathersday: 6 & 6 & 0.25 & 1 & 0 & Not a valid event \\ \hline 
		17 & -S 09-7 1:30 \newline -E 09-7 4:10& \#australia: 2, \#melbourne: 2, \#travel: 2, \#hosierlane: 1, @fedsquare: 1 & 9 & 0.25 & 0.889 & 0 & Not a valid event \\ \hline 
		18 & -S 10-27 1:40 \newline -E 10-27 4:00& @melbcentral: 1, @sephora: 1, @melbfestival: 1, @yarratrams: 1, \#arttram: 1 & 6 & 0.25 & 0.833 & 0 & Not a valid event \\ \hline 
		19 & -S 11-1 7:20 \newline -E 11-1 9:50& @fucktyler: 3, \#melbourne: 3, @converse: 3, @stacksroyal: 1, @mikel: 1 & 9 & 0.25 & 1.222 & 0 & Not a valid event \\ \hline 
		20 & -S 12-30 23:20 \newline -E 12-31 2:10& \#melbourne: 8, \#job: 3, \#culturaltart: 2, \#money: 1, @acmi: 1 & 28 & 0.25 & 0.536 & 0 & Not a valid event \\ \hline\hline 
		
	\end{tabular}
\end{table}

\textbf{\textit{Precision Evaluation.}} We select a random 20 events as the evaluation set. Table \ref{tab:clus_precision_results} shows the manual evaluation results for all events. For each event, we calculate the  the event start and end date-time by using the date-time of all tweets assigned to each cluster.  We also report area/region in $sqkm$, tweets count, top 5 hashtags/mentions with its occurrences, event description and SI index for all events. In total, 9 out of 20 are correct events according to the manual evaluation, with a precision measure of 45\%. It is clear that the precision is very low compared to our results. The reason is that using small values for $k$ and $N$ resulted in large number of detected events, where the majority of these events are just noise. This can be improved by increasing the values of $k$, and $N$ which will result in smaller number of detected events and accordingly less false positives.

\begin{table}[!th]
	\centering
	\caption{Recall results for clustering based event detection: Melbourne common events occurred in 2017}
	\small
	\label{tab:clus_recall_results}
	\setlength\tabcolsep{3pt}
	\setlength\extrarowheight{3pt}
	\begin{tabular}{cm{3cm}m{0.16\linewidth}m{0.46\linewidth}ccc}
		\hline\hline
		
		\textbf{\#} & \textbf{Event Description} & \textbf{ Date-time} & \multicolumn{1}{c}{ \textbf{Entity: occurrences }} & \textbf{*} & \textbf{Area} & \textbf{SI}  \\ \hline\hline
		1&The Night Market; Wed 6-9pm
		&-S 08-23 8:40 \newline	-E 08-23 10:20& \#melbourne: 17, \#moomba: 6, \#vscocam: 3, \#vsco: 3, \#job: 3 &65 &0.25 &0.5 \\ \hline
		
		2&Moomba Festival;  10-13 Mar 
		&-S 03-12 1:00 \newline	-E 	03-12 3:50& \#melbourne: 4,  \#davidhockney: 1, \#shrineofremembrancemelbourne: 1, \#ladiesinblack: 1, @moombafestival: 1&16&7.04&0.498\\ \hline
		
		3&Melbourne Food and Wine Festival; Mar 31 – Apr 9
		&  \multicolumn{5}{c}{ {\color{blue} \textit{Not Detected} }}\\ \hline

		4&White Night Melbourne; Feb 18
		&-S 02-18 10:30	\newline	-E 02-18 13:20& \#whitenightmelb: 3, \#whitenight: 5, \#victoria: 1, \#melbournelife: 1, \: 0 &7 &0.25 &1.43 \\ \hline

		5&Patti Smith performs; Apr 16
		& -S 04-16 10:50 \newline	-E 	04-16 13:40& \#melbourne: 5, \#music: 4, \#pattismith: 4, \#livemusic: 4, \#horses: 3 &12 &0.25 &1.67 \\ \hline

		6&Australia Day; Jan 26
		&-S 01-26 6:50 \newline	-E 	01-26 9:40& \#melbourne: 14, \#australia: 7, \#australiaday: 3, \#follow: 1, \#aperolspritz: 1 &43 &0.25 &0.6 \\ \hline
		
		7&Labour Day; Mar 13 
		&-S 03-13 2:10 \newline	-E 	03-13 3:40& \#fun: 1, \#drinks: 1, \#friends: 1, \#labourday: 1, \#rooftopbarmelbourne: 1 &5 &0.25 &1 \\ \hline
		
		8&Good Friday; Apr 14
		&-S 04-14 1:50 \newline	-E 	04-14 4:40& \#goodfridayappeal: 3, @goodfriappeal: 2, \#kidsdayout: 2, \#easter: 1, \#goodfriday: 4 &12 &0.25 &1 \\ \hline
		
		9&Easter Sunday; Apr 16
		&-S 04-16 9:00 \newline	-E 	04-16 11:50& \#melbourne: 13, \#love: 2, @localfunnyman: 2, \#easter: 1, \#hiring: 1 &51 &0.25 &0.37 \\ \hline
		
		10&ANZAC Day; Apr 25
		&	-S 04-25 4:00 \newline	-E 	04-25 6:50 & \#anzacday: 21, @mcg: 8, \#lestweforget: 8, \#afldonspies: 7, \#mcg: 8 &64 &0.25 &0.81 \\ \hline
		
		11&Queen's Birthday; Jun 12& \multicolumn{5}{c}{ {\color{blue} \textit{Not Detected} }}\\ \hline

		12&AFL Grand Final Holiday; Sep 29
		&-S 09-29 0:10 \newline	-E 	09-29 3:00& \#aflgf: 8, \#gotiges: 8, \#aflfinals: 4, \#grandfinal: 2, \#afl: 10 &22 &0.25 &1.46 \\ \hline

		13&Melbourne Cup; Nov 7
		&-S 11-7 5:00 \newline	-E 	11-7 7:50& \#melbournecup: 17, \#flemington: 6, \#melbourne: 17, \#melbcupcarnival: 2, \#love: 2 &31 &0.25 &1.42 \\ \hline
		
		14&Christmas Day; Dec 25, 2017
		&-S 12-25 0:00 \newline	-E 	12-25 2:50& \#melbourne: 2, \#christmas: 3, \#theculinaryclub: 2, \#lunchathome: 2, \#christmasday: 1 &23 &0.25 &0.43 \\ \hline
		
		15&Boxing Day; Dec 26
		&-S 12-26 20:00	 \newline -E 	12-26 22:40& \#queue: 1, @patriciacrowth3: 1, \#imagesbyhannahk: 1, \#boxingday: 1, \#fitness: 1 &13 &0.25 &0.39 \\ \hline\hline
	\end{tabular}
\end{table}

\textbf{\textit{Recall Evaluation}}. We use the same 15 common events used for recall assessment in our approach. We obtained a total of 13 correctly detected events, according to the manual evaluation, with a recall measure of 86.7\% which is higher than the recall measure of our approach. The reason is that the parameters used in this experiment result in large number of true negatives and false positives. Parameters should be tuned to balance the trade off between recall and precision. Table \ref{tab:clus_recall_results} reports the date-time, top 5 hashtags/mentions with its occurrences, tweets count, event area and SI for all events. The false negatives are highlighted in blue in the table. 

\begin{figure}[!th]
	\centering
	\includegraphics[width=.6\linewidth]{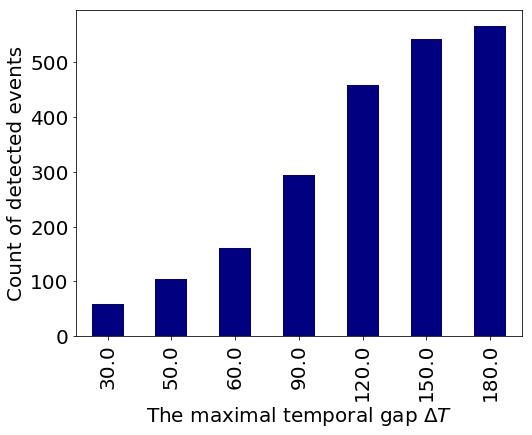}
	\caption{The number of detected events using different  values for $\Delta T$ (in minutes) }
	\label{fig:k10n15}
\end{figure}

\textbf{\textit{Parameters Effect}}. A different values for each of $\Delta T$, $K$, and $N$ have been used to their their effect on the total number of detected events as well as precision and recall measures. In total we ran 21 experiments, each has different set of parameters. For each experiment, we record the total number of significant unions/events and the detected events, as shown in Table  \ref{tab:exp2_param}. From the table, the smallest number of events in Melbourne 2017 is detected using the parameters: $\Delta T=30$minutes, $K=10$, and $N=15$, with a total of 59 events (highlighted in red in the table). While the largest number of events is detected using the parameters: $\Delta T=3$hrs, $K=2$, and $N=5$, with a total of 154546 events (highlighted in blue in the table). This shows the great impact of parameter selection on the final detected events. Figure \ref{fig:k10n15} also visualises the effect of the temporal gap parameter $\Delta T$ on the total detected events.

\begin{table*} [!th]
	
	\caption{Clustering-based event detection: The impact of parameter selection on the final detected events}
	\centering 
	\setlength\extrarowheight{1pt}
	\setlength\tabcolsep{6pt}
	\label{tab:exp2_param}
	\begin{tabular}{c|ccccc} 
		\hline
		$\Delta T$ & \multirow{2}{*}{K} & \multirow{2}{*}{N} & \# of significant & \# of unique & run time \\
		\textit{in minutes} & && unions &  events & \textit{in seconds }\\
		\hline
		\multirow{3}{*}{30} &2 &	5&	8776&	2722&	283 \\
		&5&	10&	767&	223&	244 \\
		&10&	15&		{\color{red}223}&{\color{red}	59}&	{\color{red}237 }\\
		\hline
		\multirow{3}{*}{50} &2&	5&	24517&	3020&	330\\
		&5&	10&	2730&	495&	243\\
		&10&	15&	635&	105&	233\\		
		\hline
		\multirow{3}{*}{60} & 2&	5&	33725&	3022&	364\\
		&5&	10&	4424&	607&	250\\
		&10&	15&	1000&	161&	235\\
		\hline		
		\multirow{3}{*}{90} & 2&	5&	58526&	3118&	463\\
		&5&	10	&12999&	872&	294\\
		&10&	15&	3009&	294&	250\\
		\hline		
		\multirow{3}{*}{120}& 2&	5&	78210&	3435&	545\\
		&5&	10&	25470&	1032&	349 \\
		&10	&15	&7081&	457&	274 \\
		\hline
		\multirow{3}{*}{150}&2	&5&	137564&	3735&	748\\
		&5&	10&	38364&	1119&	403\\
		&10&	15&	13585&	541&	305\\
		\hline
		\multirow{3}{*}{180}& 2	&5&	{\color{blue} 154546} &		{\color{blue}3994}&		{\color{blue}824}\\
		&5&	10	&49796	&1156	&451\\
		&10&	15&	21584&	566&	342\\
		\hline
		
	\end{tabular}
\end{table*}

We also compute the precision and recall using the parameters set that yields the smallest number of detected events, i.e. 59 events, The used parameters are: $\Delta T=30$minutes, $K=10$, and $N=15$. This is to confirm that tuning the parameters to get a smaller number of detected events will result in an increase for the precision while decrease the recall. For precision, 15 out of 20 events are correctly detected, resulting in precision measure of 75\%.
For recall, only 4 out of 15 events were correctly detected, with a recall measure of 26.7\%. The detected events belong to "Australia Day", "ANZAC Day", "AFL Grand Final Holiday", and "Boxing Day". The evaluation measures for both the clustering-based method and ours are summarized in Table \ref{tab:eval-measures}. The performed quantitative comparison shows that our method outperforms the clustering-based method.

In addition, the qualitative comparison shows that our method has an advantage over the clustering-based method, where it is able to automatically ignore tweets that do not belong to any event. Accordingly it can be effectively used with the online stream without any topic or keyword filtering. While the clustering-based method assigns each tweet to a cluster. This means that a set of non-event tweets with small temporal and spatial proximity will eventually form a significant cluster. This limits the application of the clustering method to the tweets with specific keywords in order to obtain the desirable results, as the authors of the paper did in the twitter case study where they  applied the algorithm on the weather-related tweets dataset.

\begin{table*}[!t]
	
	\caption{Evaluation measures for event detection using our method and the clustering baseline method}
	\centering 
	
	\setlength\extrarowheight{2pt}
	\setlength\tabcolsep{6pt}
	\label{tab:eval-measures}
	\begin{tabular}{llcc} 
		\hline
		Method & Parameters &Precision& Recall \\
		\hline
		Clustering method &   $\Delta T=30$minutes, $K=10$, and $N=15$&75\%&26.7\% \\
		Clustering method &  $\Delta T=3$hrs, $K=2$, and $N=5$ &45\%&86.7\%\\
		Our Method & $\Delta T=10$minutes, $\tau_1=0.01$, $\tau_2=0.4$  &89\%&66.7\% \\
		\hline
	\end{tabular}
\end{table*}

\section*{Conclusion}
\label{sectConclusion}
In this paper, we present a multiscale spatio-temporal real-time event detection approach which is capable of detecting social media events of different spatial and temporal resolution in real-time. 
This approach utilizes a quad-tree data structure to identify events with varying spatial coverage, and a Poisson model with a smoothing function to detect previously unseen events with different temporal resolutions.
Also, the proposed method is an unsupervised approach that does not require a list of defined topics for event detection and effectively detects both local and global events. The method is evaluated using different social media datasets: Twitter and Flickr; for different cities: Melbourne, London, Paris and New York. The experiments have demonstrated that the proposed method achieves better results than the baseline algorithm. 

In the future, we plan to improve our method by taking into account the changing structure of the constructed quad-tree over time. Also, more experiments will be conducted to fine-tune the parameters of the proposed method using different datasets. The proposed method will be extended to use non-geotagged social media data based on textual information~\cite{Han-jair14,chi-wnut16,lim-iui19,Rahimi-acl17,Chong-tois18}. Finally, we can also improve tour recommendation works by planning itineraries that avoid detected events such as accidents~\cite{chen-its14,friggstad2018orienteering,padia-bigdata19,liebig-is17,liebig-mud14}.



\begin{backmatter}

\section*{Ethics approval and consent to participate}
Not applicable.

\section*{Consent for publication}
The authors consent to the publication of this manuscript.

\section*{Availability of data and materials}
Not applicable.

\section*{Competing interests}
The authors declare that they have no competing interests.

\section*{Funding}
This research is funded by MyIP:7293 and SRG-ISTD-2018-140.

\section*{Author's contributions}
YG designed the research, ran experiments, analyzed results, and wrote the manuscript. SK, AH and KHL designed the research, analyzed results and contributed to manuscript preparation. All authors read and approved the final manuscript.

\section*{Acknowledgements}
This research is funded in part by the Defence Science and Technology Group, Edinburgh, South Australia, under contract MyIP:7293, and the Singapore University of Technology and Design under grant SRG-ISTD-2018-140.

\bibliographystyle{bmc-mathphys} 
\bibliography{locEventDetect}      


\end{backmatter}
\end{document}